\documentclass[aps,nofootinbib,superscriptaddress,showpacs,floatfix,preprintnumbers,twocolumn]{revtex4}
\usepackage{amssymb,amsmath,graphicx,xcolor,slashed}
\usepackage[colorlinks=true,linkcolor=blue,citecolor=blue,urlcolor=blue]{hyperref}
\usepackage{cleveref}
\usepackage{physics}
\usepackage{subfigure}
\usepackage{multirow}
\usepackage{longtable}
\usepackage{pdflscape}
\usepackage{soul,xcolor}
\setstcolor{red}

\newcommand{\non}{\nonumber}

\newcommand{\LambdaQCD}{\Lambda_\text{QCD}}
\newcommand{\diff}[2]{\operatorname{d}{\hspace{-0.15em}}^{#1}{#2}\hspace{0.15em}}
\newcommand{\MSbar}{\overline{\text{MS}}}
\newcommand{\N}{NLO}
\newcommand{\NL}{NLO+LRR}
\newcommand{\NR}{NLO$\times$RGR}
\newcommand{\NLR}{(NLO+LRR)$\times$RGR}
\newcommand{\NN}{NNLO}
\newcommand{\NNL}{NNLO+LRR}
\newcommand{\NNR}{NNLO$\times$RGR}
\newcommand{\NNLR}{(NNLO+LRR)$\times$RGR}

\begin{document}
\title{Systematic Improvement of $x$-dependent Unpolarized Nucleon Generalized Parton Distribution in Lattice-QCD Calculation}

\author{Jack Holligan}
\email{holligan@msu.edu}
\author{Huey-Wen Lin}
\email{hwlin@pa.msu.edu}
\affiliation{Department of Physics and Astronomy, Michigan State University, East Lansing, MI 48824}

\preprint{MSUHEP-23-033}

\pacs{12.38.-t, 
      11.15.Ha,  
      12.38.Gc  
}

\begin{abstract}
We present a first study of the effects of renormalization-group resummation (RGR) and leading-renormalon resummation (LRR) on the systematic errors of the unpolarized isovector nucleon generalized parton distribution in the framework of large-momentum effective theory (LaMET).
This work is done using lattice gauge ensembles generated by the MILC collaboration, consisting of 2+1+1 flavors of highly improved staggered quarks with a physical pion mass at lattice spacing $a\approx 0.09$~fm and a box width $L\approx 5.76$~fm.
We present results for the nucleon $H$ and $E$ GPDs with average boost momentum $P_z\approx 2$~GeV at momentum transfers $Q^2=[0, 0.97]$~GeV$^2$ at skewness $\xi=0$ as well as $Q^2\in 0.23$ GeV$^2$ at $\xi=0.1$, renormalized in the $\overline{\rm MS}$ scheme at scale $\mu=2.0$ GeV, with two- and one-loop matching, respectively.
We demonstrate that the simultaneous application of RGR and LRR significantly reduces the systematic errors in renormalized matrix elements and distributions for both the zero and nonzero skewness GPDs, and that it is necessary to include both RGR and LRR at higher orders in the matching and renormalization processes.
\end{abstract}

\maketitle

\section{Introduction}
An open question in the theory of quantum chromodynamics (QCD) is how the fundamental degrees of freedom, quarks and gluons, comprise the more massive hadrons.
The quarks and gluons (known collectively as partons) contribute to a hadron's mass and spin but cannot be studied in isolation due to confinement.
Thus, knowledge of the internal structure of a hadron is highly valued.
Great effort has been focused on the study of parton distribution functions (PDFs), which describe the distribution of a hadron's longitudinal momentum among its constituents, and much has been learned about hadronic structure from these studies (see Ref.~\cite{Amoroso:2022eow} for a review from Snowmass~2021).
However, the PDF only paints a one-dimensional picture of the hadron, since it is dependent solely on longitudinal momentum.
Generalized parton distributions (GPDs) contain more information about the hadron, including spin structure, form factors and how the longitudinal momentum of the parton depends on the distance from the center of the hadron.
The unpolarized GPD is comprised of two functions commonly denoted $H$ and $E$, defined in terms of matrix elements on the lightcone as
\begin{widetext}
\begin{align}\label{eq.LightconeGPD}
    F(x,Q^2,\xi) &= \int\frac{\diff{}{z^-}}{4\pi} e^{ixp^+z^-} \left\langle p''\left|\overline{\psi}\left(-\frac{z}{2}\right)\gamma^+ \mathcal{L}\left(-\frac{z}{2},\frac{z}{2}\right)
    \psi\left(\frac{z}{2}\right) \right|p'\right\rangle\non\\
    &=\frac{1}{2p^+}\left[H(x,Q^2,\xi)\overline{u}(p'')\gamma^+u(p')+E(x,Q^2,\xi)\overline{u}(p'')\frac{i\sigma^{+\nu}Q_{\nu}}{2m}u(p')\right],
\end{align}
\end{widetext}
where $\mathcal{L}(-z/2,z/2)$ is a link along the lightcone,
$Q^\mu=(p''-p')^\mu$ is the momentum transfer,
and $\xi=\frac{p''^+ - p'^+}{p''^+ + p'^+}$ is the skewness.
In the limit $Q^2 \to 0$ and $\xi \to 0$, the $H$ GPD reduces to the PDF.
The $E$ GPD is inaccessible in this limit, since it is multiplied by the momentum transfer vector.
GPDs can be probed experimentally by processes such as deeply-virtual Compton scattering (DVCS)~\cite{Ji:1996nm,Braun:2022bpn} or deeply-virtual meson production (DVMP)~\cite{Kriesten:2019jep}, and their study will be an important experimental program at the future Electron-Ion Collider (EIC)~\cite{AbdulKhalek:2021gbh,Achenbach:2023pba,Abir:2023fpo,AbdulKhalek:2022hcn,Burkert:2022hjz}.

Lattice QCD involves converting the QCD path integral from continuous Minkowski spacetime to discrete Euclidean spacetime, making field-theory calculations amenable to supercomputers.
It can provide early insight into GPD functions complementary to experimental programs.
The computation of the Bjorken-$x$ dependence of parton distributions can be studied in the framework of lattice QCD using one of several recent methods:
the ``hadronic-tensor approach''~\cite{Liu:1993cv,Liu:1998um,Liu:1999ak,Liu:2016djw,Liu:2017lpe,Liu:2020okp},
the Compton-amplitude approach (or ``OPE without OPE'')~\cite{Aglietti:1998ur,Martinelli:1998hz,Dawson:1997ic,Capitani:1998fe,Capitani:1999fm,Ji:2001wha,Detmold:2005gg,Braun:2007wv,Chambers:2017dov,Detmold:2018kwu,QCDSF-UKQCD-CSSM:2020tbz,Horsley:2020ltc,Detmold:2021uru},
the ``current-current correlator'' method~\cite{Braun:2007wv,Ma:2017pxb,Bali:2017gfr,Bali:2018spj,Joo:2020spy,Gao:2020ito,Sufian:2019bol,Sufian:2020vzb}, or large-momentum effective theory (LaMET)~\cite{Ji:2013dva,Ji:2014gla,Ji:2020ect}, which is our focus in this paper.

The method of LaMET begins with the study of spatially separated, equal time, matrix elements of boosted hadrons computed directly on the lattice:
\begin{equation}\label{eq:hB}
    h^B(z,Q^2,\xi)=\mel{p''}{\overline{\psi}\left(-\frac{z}{2}\right)\Gamma W\left(-\frac{z}{2},\frac{z}{2}\right)\psi\left(\frac{z}{2}\right)}{p'}
\end{equation}
where $\Gamma=\gamma_t,\,\gamma_t\gamma_5,\,\gamma_t\gamma_{
\perp}$ for the unpolarized, helicity and transversity GPDs, respectively.
$W(-z/2,z/2)$ is a lattice link from the coordinate $(0,0,0,-z/2)$ to $(0,0,0,z/2)$, since we may assume without loss of generality that the average momentum $(\vec{p'}+\vec{p''})/2$ is along the $z$ axis.
The bare matrix elements are then renormalized and Fourier transformed to momentum space to obtain the quasi-GPD.
The final step is to match the quasi-GPD to the lightcone to obtain the GPD.
GPDs have been studied in LaMET in the Breit-frame setting on the lattice.
The GPD on the lattice was first studied in the case of the pion in Ref.~\cite{Chen:2019lcm} and carried out at physical pion mass~\cite{Lin:2023gxz,Lin:2023kxn} by MSULat in the zero-skewness limit.
The nucleon unpolarized and helicity GPDs were studied in Refs.~\cite{Lin:2020rxa,Lin:2021brq,Alexandrou:2020zbe} and the transversity ones in Ref.~\cite{Alexandrou:2021bbo}.
Recently, the ETMC and ANL/BNL groups have computed bare matrix elements in asymmetric frames~\cite{Bhattacharya:2022aob} to help reduce the computational cost of the lattice calculation.

Since the aforementioned numerical studies of GPDs, developments in the framework of LaMET include renormalization-group resummation (RGR)~\cite{Su:2022fiu} and leading-renormalon resummation (LRR)~\cite{Zhang:2023bxs}.
RGR is designed to resum the logarithms that arise from the differing intrinsic physical scale and final renormalization scale of the parton.
The method is to set the energy scale such that the logarithmic terms vanish and then evolve to the desired scale with the renormalization group.
This process can be applied both to the renormalization of the bare matrix elements as well as the perturbative matching.
LRR is designed to resum the divergence arising from the infrared renormalon (IRR) which plagues perturbation series~\cite{Zichichi:1979gj}, and whose effect is more pronounced with the application of RGR alone.
The first application of LRR was to the pion PDF in Ref.~\cite{Zhang:2023bxs}, which showed that LRR in combination with RGR results in greatly reduced systematic uncertainties in the final $x$-dependent PDF.
The ANL/BNL collaboration also applied LRR (and RGR) to their LaMET calculation of the nucleon transversity PDF in Ref.~\cite{Gao:2023ktu} to better control the systematic errors.
The field of LaMET has matured to the point at which such systematic uncertainties become an important issue.
The methods of RGR and LRR have not yet been applied to GPDs;
doing so can lead to more precise calculation of tomography from lattice QCD in the future.

The purpose of this paper is to make the first application of the RGR and LRR improvements to the calculation of the unpolarized nucleon isovector GPD at different skewness, $\xi$, and squared momentum transfer, $Q^2$, in the Breit frame.
We use clover valence fermions at physical quark mass with a lattice spacing of $a\approx 0.09$~fm and box length $L=64a \approx 5.76$~fm with QCD vacuum composed of $N_f=2+1+1$ flavors of highly-improved staggered quarks~\cite{Follana:2006rc}, generated by the MILC collaboration~\cite{MILC:2010pul,MILC:2012znn,MILC:2015tqx} with one step of hypercubic smearing~\cite{Hasenfratz:2001hp} applied to the gauge links to reduce discretization effects.
The valence fermion parameters are tuned so as to produce a physical pion mass ($m_{\pi}\approx 130$~MeV).
The same mixed-action setup used in this calculation was previously studied in Refs.~\cite{Mondal:2020cmt,Park:2020axe,Jang:2019jkn,Jang:2019vkm,Gupta:2018lvp,Lin:2018obj,Gupta:2018qil,Gupta:2017dwj,Bhattacharya:2015wna,Bhattacharya:2015esa,Bhattacharya:2013ehc,Bhattacharya:2011qm,Briceno:2012wt,Yoon:2016jzj,Bhattacharya:2016zcn} and found to be free of exceptional configurations which can cause the Dirac matrix to be ill-conditioned or the correlation functions to be anomalously large.
From a total of 1960 lattice configurations, we use the 501,760 measurements of the bare nucleon matrix elements of Eq.~\ref{eq:hB} with average boost momentum $P_z=10\times\frac{2\pi}{L}\approx 2.2$~GeV in Ref.~\cite{Lin:2020rxa}. 
More information on the bare matrix elements such as the source-sink separation, the momentum smearing and momentum transfer can be found in Ref.~\cite{Lin:2020rxa} and its supplemental material.
The ground-state nucleon bare matrix elements are extracted by simultaneously fitting multiple source-sink separations with skewness values of $\xi=0$ and $\xi=0.1$.
For each skewness value, we have momentum transfer $Q^2\in\{0.0,0.19,0.39,0.77,0.97\}$~GeV$^2$ and $Q^2=0.23$~GeV$^2$ respectively.

This paper is laid out as follows.
In Sec.~\ref{sec:xi0}, we describe the methodology of RGR and LRR as well as the outline of our calculation of the GPDs at zero skewness from the bare matrix elements.
We also show results for zero-skewness GPDs for both zero and nonzero momentum transfer, demonstrating the improvements afforded by both RGR and LRR as well as matching at both next-to-leading-order (NLO) and next-to-next-to-leading-order (NNLO).
In Sec.~\ref{sec:xinonzero}, we show nonzero-skewness GPDs at \N.
We conclude in Sec.~\ref{sec:Conclusion}.

\section{Zero-Skewness GPDs at NLO and NNLO}\label{sec:xi0}

In this section we present the zero-skewness ($\xi=0$) unpolarized isovector nucleon GPD at both zero ($H$ GPD only) and nonzero ($H$ and $E$ GPDs) momentum transfer $Q^2$.
When both $\xi=0$ and $Q^2=0$, the unpolarized GPD reduces to the unpolarized PDF.
The renormalization procedure and the transformation to momentum space are also described in this section, since the same methods are used for all values of momentum transfer and skewness.
We describe the lightcone matching for the case of zero skewness and postpone the discussion of nonzero skewness matching to Sec.~\ref{sec:xinonzero}.

\begin{figure*}[htp]
\centering
\subfigure{\includegraphics[width=0.3\linewidth]{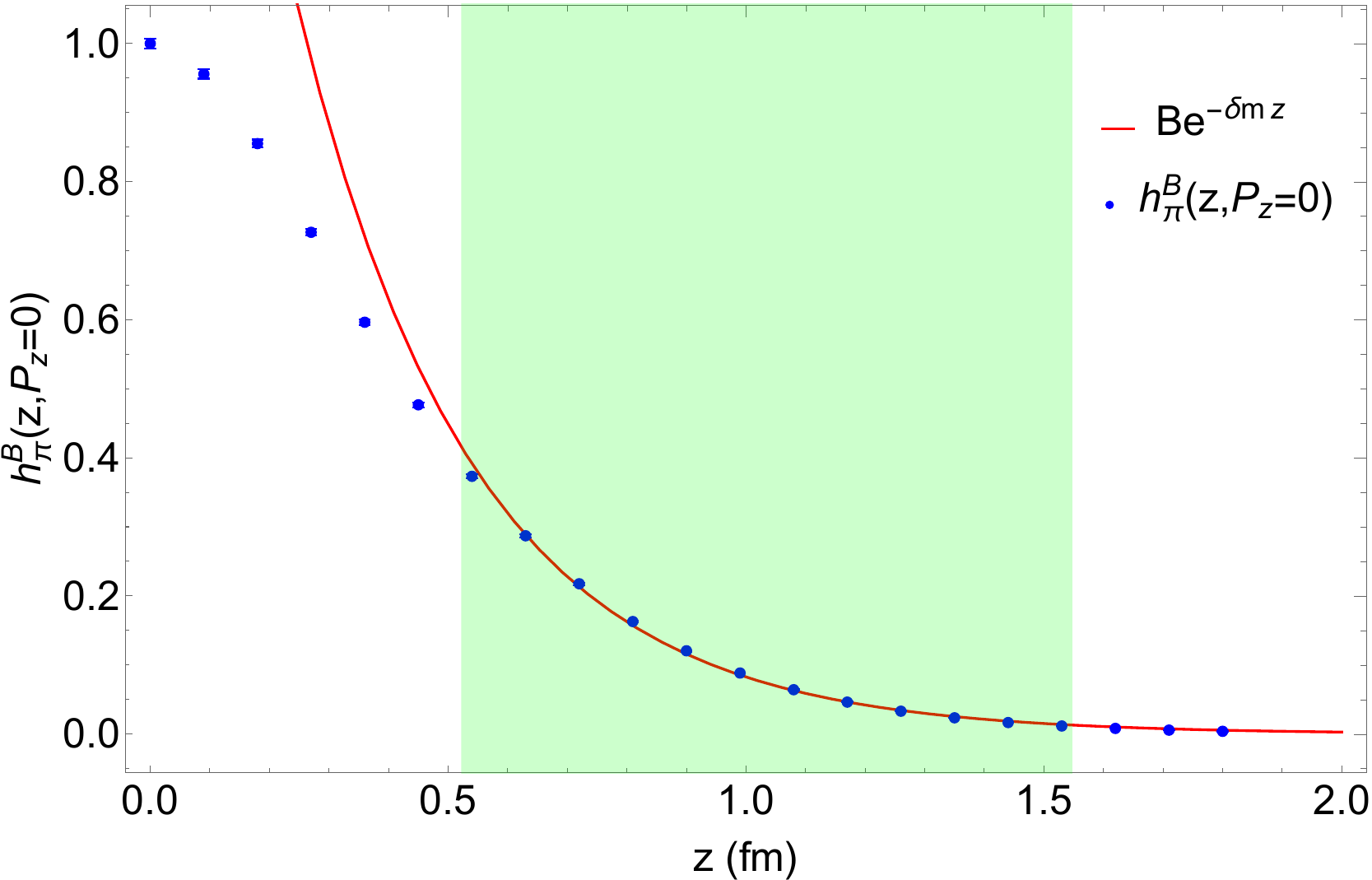}}\quad
\subfigure{\includegraphics[width=0.3\linewidth]{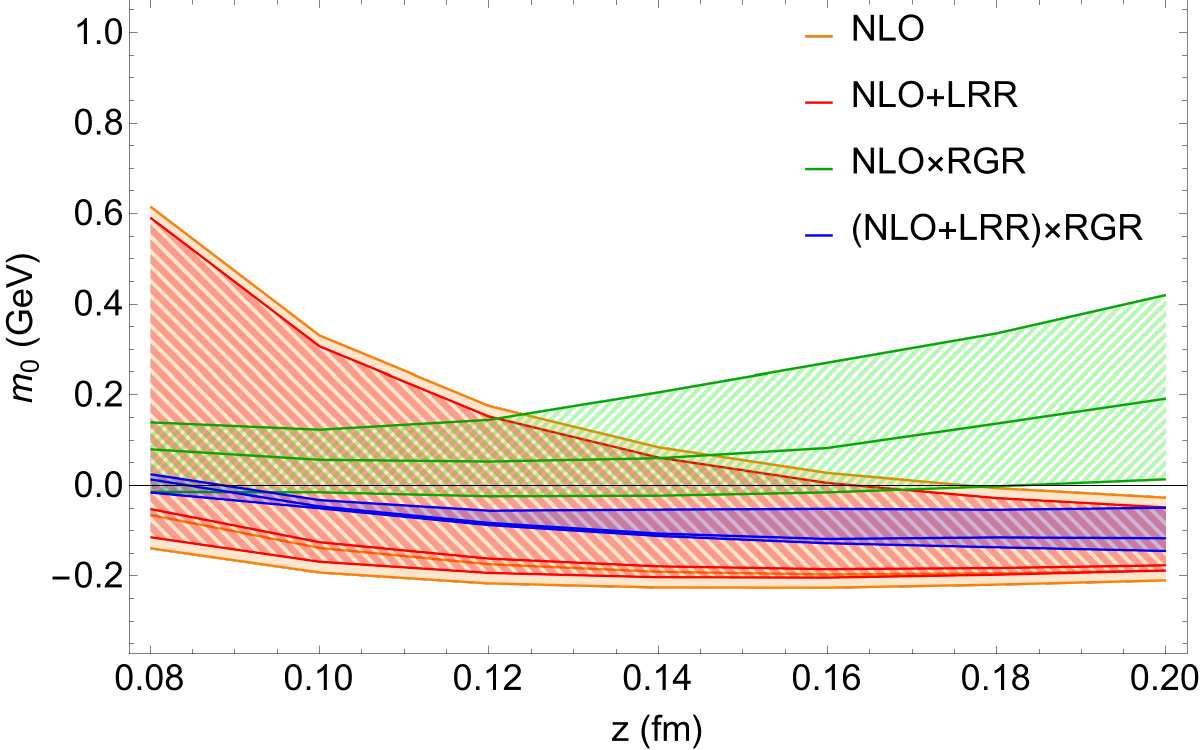}}\quad
\subfigure{\includegraphics[width=0.3\linewidth]{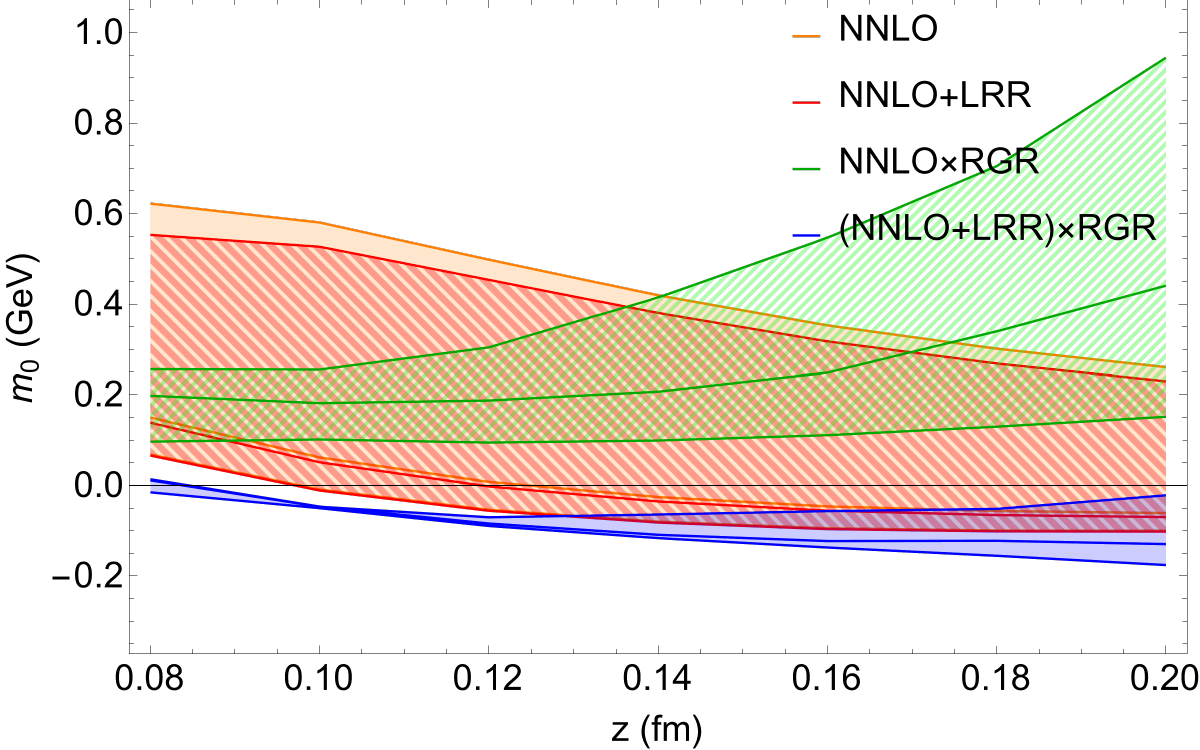}}
\caption{Determination of the linear divergence (left-most plot), $\delta m$, by fitting the zero-momentum pion matrix element (blue points) to the function $Be^{-\delta m\,z}$ (red curve) in the interval $z=[0.54,1.53]$ fm (shaded green). The error bars for the pion matrix elements are included but too small to be visible. The middle (right-most) plots show the renormalon divergence, $m_0$, determined to (N)\N~(solid orange), (N)\NL~(hatched red), (N)\NR~(hatched green) and ((N)NLO+LRR)$\times$RGR~(solid blue). The vertical width of each band corresponds to the systematic error determined from scale variation described in Sec.~\ref{subsec:Q0xi0}.}
\label{fig:linDivm0Plots}
\end{figure*}

We begin with the renormalization of the bare matrix elements.
We perform the renormalization in the hybrid-ratio scheme~\cite{Ji:2020brr}, in which the bare matrix elements are renormalized in the ratio scheme up to distances $z_s=3a \approx 0.27$~fm with our lattice spacing, and at large distances the linear divergence and renormalon divergence are removed by an exponential term.
The ratio scheme involves dividing the bare matrix element at nonzero boost momentum by those at zero boost momentum at fixed $z$.
The fully renormalized matrix element (for both $H$ and $E$) is given by
\begin{equation}\label{eq:hR}
  h^\text{R}(z,P_z,Q^2,\xi) = \begin{cases}
	\frac{h^\text{B}(z,P_z,Q^2,\xi)}{h^\text{B}_{\pi}(z,P_z=0)} & z < z_s \\
	e^{(\delta m+m_0)(z-z_s)} \frac{h^\text{B}(z,P_z,Q^2,\xi)}{h^\text{B}_{\pi}(z_s,P_z=0)} & z \geqslant z_s
  \end{cases}
\end{equation}
where we have used bare unpolarized pion matrix elements at zero boost-momentum, $h^\text{B}_{\pi}(z,P_z=0)$~\cite{LatticePartonCollaborationLPC:2021xdx} for the ratio scheme at $z<z_s$.
At $Q^2=0$, we normalize the matrix elements to 1 at $z=0$.
The terms $\delta m$ and $m_0$ are, respectively, the linear divergence and the renormalon divergence.
The linear divergence is due to the self-energy of the Wilson line in the bare matrix element, and the renormalon divergence arises from the fact that the perturbation series used to calculate $\delta m$ is not convergent to all orders~\cite{LatticePartonCollaborationLPC:2021xdx,Ji:2020brr,Zhang:2023bxs,Zichichi:1979gj}.
We determine the linear divergence by following the same procedure as in Ref.~\cite{Ji:2020brr} by fitting the zero-momentum pion matrix elements to the exponential decay $Be^{-\delta m\,z}$ in the interval $z=[0.54,1.53]$~fm, as shown in the left-most panel of Fig.~\ref{fig:linDivm0Plots}, where $B$ and $\delta m$ are fitting parameters. This same procedure was performed with the same data in our previous work \cite{PionPaper} in which we find $\delta m=0.668(10)$~GeV.

While the computation of the linear divergence would seem to be subjective, it is compensated for by the calculation of the renormalon divergence such that their sum, $\delta m+m_0$, is constant in a fixed scheme~\cite{Ji:2020brr}.
The renormalon divergence is determined by demanding that the short-distance physics ($z\lesssim 0.3$~fm) agrees with the theoretical predictions of the operator-product expansion (OPE).
The functions that appear in the OPE (and describe the short distance physics) are known as Wilson coefficients, which we denote by $C_0(z,\mu)$, where $z$ is the Wilson length, and $\mu$ is the energy scale.
For a Wilson length $z$ and renormalization scale $\mu$, which is the final desired energy scale for the lightcone PDF renormalized in the modified minimal-subtraction ($\MSbar$) scheme, the unpolarized Wilson coefficients are
\begin{equation}
    C_0^\text{NLO}(z,\mu) = 1 + \frac{\alpha_s(\mu)C_F}{2\pi}\left(\frac{3}{2}l(z,\mu) + \frac{5}{2}\right)\label{eq.CNLO}
\end{equation}
at NLO~\cite{Izubuchi:2018srq}
and
\begin{multline}
    C_0^\text{NNLO}(z,\mu) = C_0^\text{NLO}(z,\mu) + \left(\frac{\alpha_s(\mu)}{2\pi}\right)^2 \\
    \times \Bigg[l^2(z,\mu) \left(\frac{15}{2} - \frac{n_f}{3}\right)
    + l(z,\mu) \left(37.1731-\frac{5}{3}n_f\right) \\
    - 4.34259 n_f + 51.836 \Bigg]\label{eq.CNNLO}
\end{multline}
at NNLO~\cite{Li:2020xml}
where $l(z,\mu) = \ln\left(z^2\mu^2e^{2\gamma_E}/4\right)$, $\gamma_E$ is the Euler-Mascheroni constant,
$\alpha_s(\mu)$ is the strong coupling at energy scale $\mu$,
$C_F$ is the quadratic Casimir for the fundamental representation of SU(3)
and $n_f$ is the number of fermion flavors.

The determination of the renormalon divergence can be improved with the additions of renormalization-group resummation (RGR)~\cite{Su:2022fiu,Zhang:2023bxs} and leading-renormalon resummation (LRR)~\cite{Zhang:2023bxs}.
The difference between the intrinsic physical scale and the final renormalization scale results in logarithmic terms that require resummation.
This is achieved by setting the renormalization scale such that the logarithmic terms vanish and then evolving to the desired scale using the renormalization-group equation:
\begin{equation}\label{eq.RGE}
    \frac{\diff{}{C_0(z,\mu)}}{\diff{}{\ln(\mu^2)}}=\gamma(\mu)C_0(z,\mu)
\end{equation}
where $\gamma(\mu)$ is the anomalous dimension, which has been calculated up to three loops~\cite{Braun:2020ymy}.
The energy scale at which the logarithms vanish is $\mu=2e^{-\gamma_E}/z\equiv\mathtt{z}^{-1}$ as can be seen in Eqs.~\ref{eq.CNLO} and \ref{eq.CNNLO}.
Thus, we can improve the computation of the Wilson coefficient with RGR giving
\begin{align}\label{eq.CRGR}
    C_0^{\text{N}^{k}\text{LO}\times\text{RGR}}(z,\mu)&=C_0^{\text{N}^{k}\text{LO}}(z,\mathtt{z}^{-1})\nonumber\\
    &\times\exp\left(\int^{\alpha_s(\mu)}_{\alpha_s(\mathtt{z}^{-1})}\diff{}{\alpha'}\frac{\gamma(\alpha')}{\beta(\alpha')}\right)
\end{align}
where $k=1$ for NLO, $k=2$ for NNLO and $\beta(\alpha)$ is the QCD beta-function. For brevity, we define
\begin{equation}
    \mathcal{I}(\mu,\mathtt{z}^{-1})=\exp\left(\int^{\alpha_s(\mu)}_{\alpha_s(\mathtt{z}^{-1})}\diff{}{\alpha'}\frac{\gamma(\alpha')}{\beta(\alpha')}\right).
\end{equation}

The Wilson coefficients are perturbation series which can contain a renormalon divergence~\cite{Fischer:1999qm}. We account for this using the LRR method, in which the Wilson coefficient is modified according to Eq.~14 of Ref.~\cite{Zhang:2023bxs}
\begin{multline}\label{eq.CLRR}
    C^{\text{N}^{k}\text{LO+LRR}}_0(z,\mu) = C^{\text{N}^{k}\text{LO}}_0(z,\mu) \\
    +z\mu\left(C_\text{PV}(z,\mu)-\sum_{i=0}^{k-1}\alpha^{i+1}_s(\mu)r_i\right),
\end{multline}
where $r_i$ are the coefficients of the renormalon series in $\alpha_s$ and $C_\text{PV}(z,\mu)$ is the contribution of the renormalon to the Wilson coefficients after a Borel transformation originally derived in Refs.~\cite{Bali:2013pla,Pineda:2001zq}.
Explicit definitions can be found in Eqs.~12 and 13, respectively, of Ref.~\cite{Zhang:2023bxs}

We can then combine the RGR and LRR improvements into a single high-quality Wilson coefficient
\begin{multline}\label{eq.CRGRLRR}
    C^{\text{(N}^{k}\text{LO+LRR)}\times\text{RGR}}_0(z,\mu) = C^{\text{N}^{k}\text{LO+LRR}}_0(z,\mathtt{z}^{-1})\\
    \times\mathcal{I}(\mu,\mathtt{z}^{-1}).
\end{multline}
The Wilson coefficients with different improvements yield different central values and uncertainties for the renormalon divergence, $m_0$.
We use the same procedure to compute the renormalon divergence as Ref.~\cite{Zhang:2023bxs} in which $\ln\left(\frac{e^{-\delta m\,z}C_0(z,\mu)}{h^{\rm B}_{\pi}(z,P_z)}\right)$ is fitted to $m_0z+c$ for multiple sets of $z$ values. 
We interpolate the matrix elements $h^B_{\pi}(z,P_z=0)$ as in our previous work~\cite{PionPaper} and determine $m_0(z)$ with the inputs of $\{z-0.02\text{ fm},\,z,\,z+0.02\text{ fm}\}$ to a maximum of $z=0.2$~fm. 
A plot of $m_0$ to different orders and with different improvements as a function of fitting range is shown in the middle and right panels of Fig.~\ref{fig:linDivm0Plots}.
We seek a plateau in the values of $m_0$ across different fitting ranges which signals a stable and reliable measurement of the renormalon divergence and select the corresponding value as the measurement of $m_0$.
The results with the smallest errors as well as clear plateaux are those for which RGR and LRR are applied simultaneously.
Having determined both the linear divergence and the renormalon divergence, we now have fully renormalized matrix elements in the hybrid-ratio scheme (using Eq.~\ref{eq:hR}).

To obtain the quasi-distribution, we first extrapolate the renormalized matrix elements to infinite distance with a view to performing a Fourier transform.
The extrapolation model is inspired by the small-$x$ physics we expect to see in the PDF~\cite{Gao:2021dbh,Gao:2022uhg,Ji:2020brr}, which is itself governed by the large-distance behavior of the renormalized matrix elements:
\begin{equation}\label{eq.Extrapolation}
    h^\text{R}(z,Q^2,\xi) \to \frac{Ae^{-mz}}{|zP_z|^d} \quad\text{as } z\to\infty,
\end{equation}
where $A$, $m$ and $d$ are fitting parameters.
The data used to fit the extrapolation must be at sufficiently large $z$ that we can realistically model the large-distance behavior.
We then Fourier transform to momentum space to obtain the quasi-GPDs with the convention
\begin{equation}\label{eq.FourierTransform}
    qF(x,Q^2,\xi) = \int^{\infty}_{-\infty}\frac{P_z\diff{}{z}}{2\pi}e^{ixzP_z}h^\text{R}_F(z,Q^2,\xi),
\end{equation}
where $F$ is either $H$ or $E$ corresponding to the respective GPD functions.
By extrapolating the renormalized matrix elements to infinite distance, we remove unphysical oscillations from the quasi-GPDs that would otherwise occur in the Fourier transform.

The final stage in the calculation is the perturbative matching to align the ultraviolet (UV) behavior of the quasi-GPD with the lightcone.
The matching formula is
\begin{multline}\label{eq.Matchingxi0}
	qF(x,Q^2,\xi) = \int^{1}_{-1}\frac{\diff{}{y}}{|y|}\mathcal{K}(x,y,\mu,\xi,P_z)F(y,Q^2,\xi)\\
	+\mathcal{O}\left(\frac{\LambdaQCD^2}{P_z^2x^2(1-x)}\right)
\end{multline}
where $\mathcal{K}$ is the matching kernel.
Once again, this formula applies to both the quasi-$H$ and quasi-$E$ GPDs.
For zero skewness, $\xi=0$, the kernel has been calculated up to NNLO in the hybrid-ratio scheme for unpolarized GPDs in Refs.~\cite{Chen:2020ody,Li:2020xml,Su:2022fiu}.
For nonzero skewness, the kernel has been computed up to NLO for unpolarized GPDs (as well as helicity and transversity GPDs)~\cite{Yao:2022vtp},
and we discuss it in more detail in Sec.~\ref{sec:xinonzero}.

The RGR process applied to the matching is designed to resum logarithmic terms that occur in the matching kernel. The philosophy is the same as that of the determination of the renormalon divergence with RGR in that we set an energy scale such that the logarithms vanish and then evolve to the final desired energy scale. This time, the anomalous dimension is the Dokshitzer-Gribov-Lipatov-Altarelli-Parisi (DGLAP) equation
\begin{equation}\label{eq.DGLAP}
  \frac{\diff{}{F(x,\xi=0,\mu)}}{\diff{}{\ln(\mu^2)}} =
  \int^1_x \frac{\diff{}{z}}{|z|} \mathcal{P}(z) F\left(\frac{x}{z},\xi=0,\mu\right),
\end{equation}
where $\mathcal{P}(z)$ is the DGLAP kernel, which has been calculated up to three loops~\cite{Moch:2004pa}.
The formula is applicable to both $H$ and $E$ GPDs.
We use the same algorithm for RGR matching as in Ref.~\cite{Su:2022fiu}.
However, this formula is only applicable to zero-skewness GPDs.
At nonzero skewness, a different evolution formula is required for $|x|<\xi$. The corresponding formula is the Efremov-Radyushkin-Brodsky-Lepage (ERBL) equation~\cite{Efremov:1978rn,Efremov:1979qk,Lepage:1979zb,Lepage:1980fj} and $|x|<\xi$ is known as the ERBL region.
In this $x$ range there are two distinct scales that emerge, which cannot be eliminated simultaneously by the choice of a single energy scale.
A more sophisticated technique must be developed for this case in the future.

\subsection{Zero-Skewness $H$ GPD at $Q^2=0$}\label{subsec:Q0xi0}
\begin{figure*}[htp]
  \centering
  \subfigure{\includegraphics[width=0.4\linewidth]{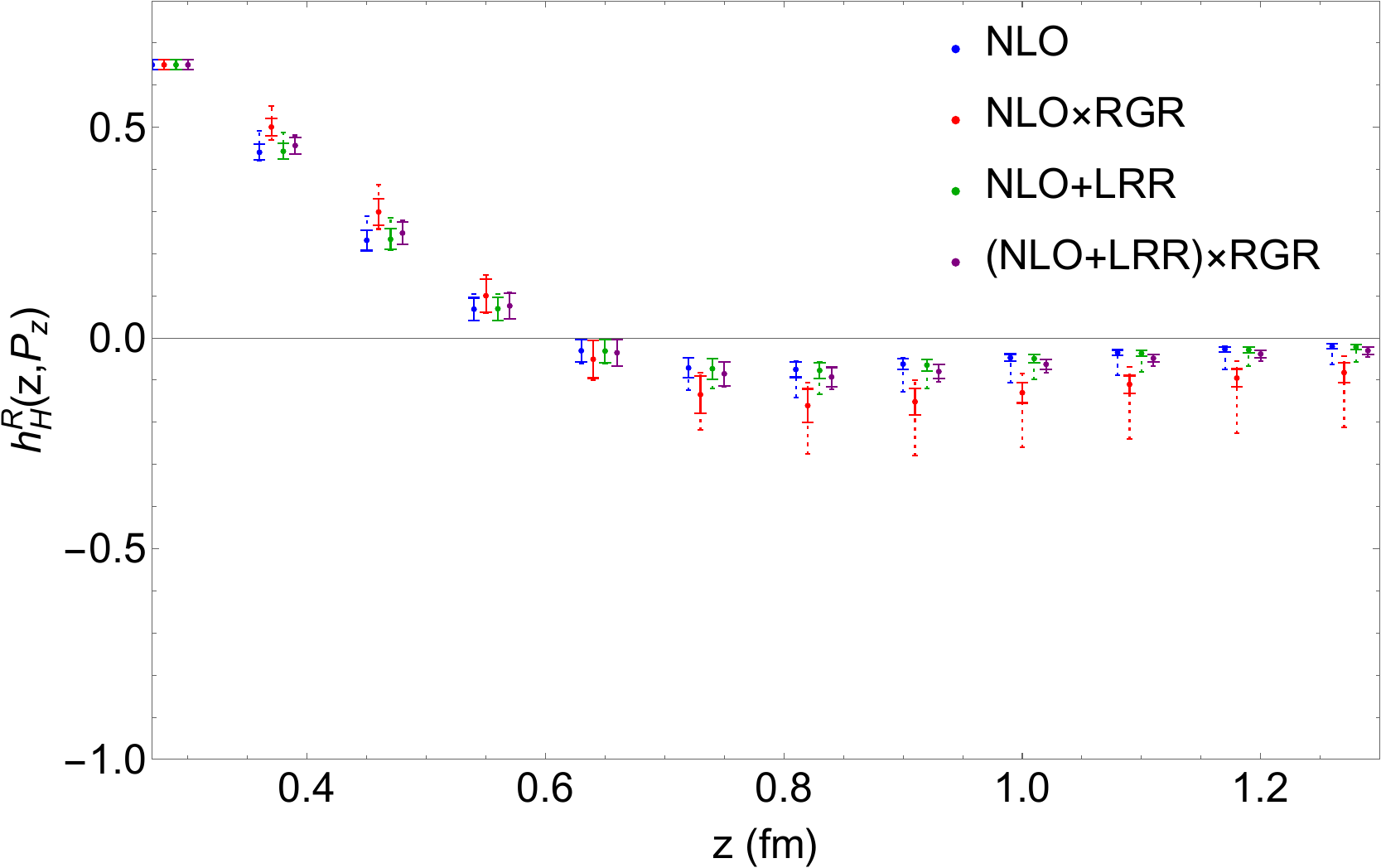}}\quad
  \subfigure{\includegraphics[width=0.4\linewidth]{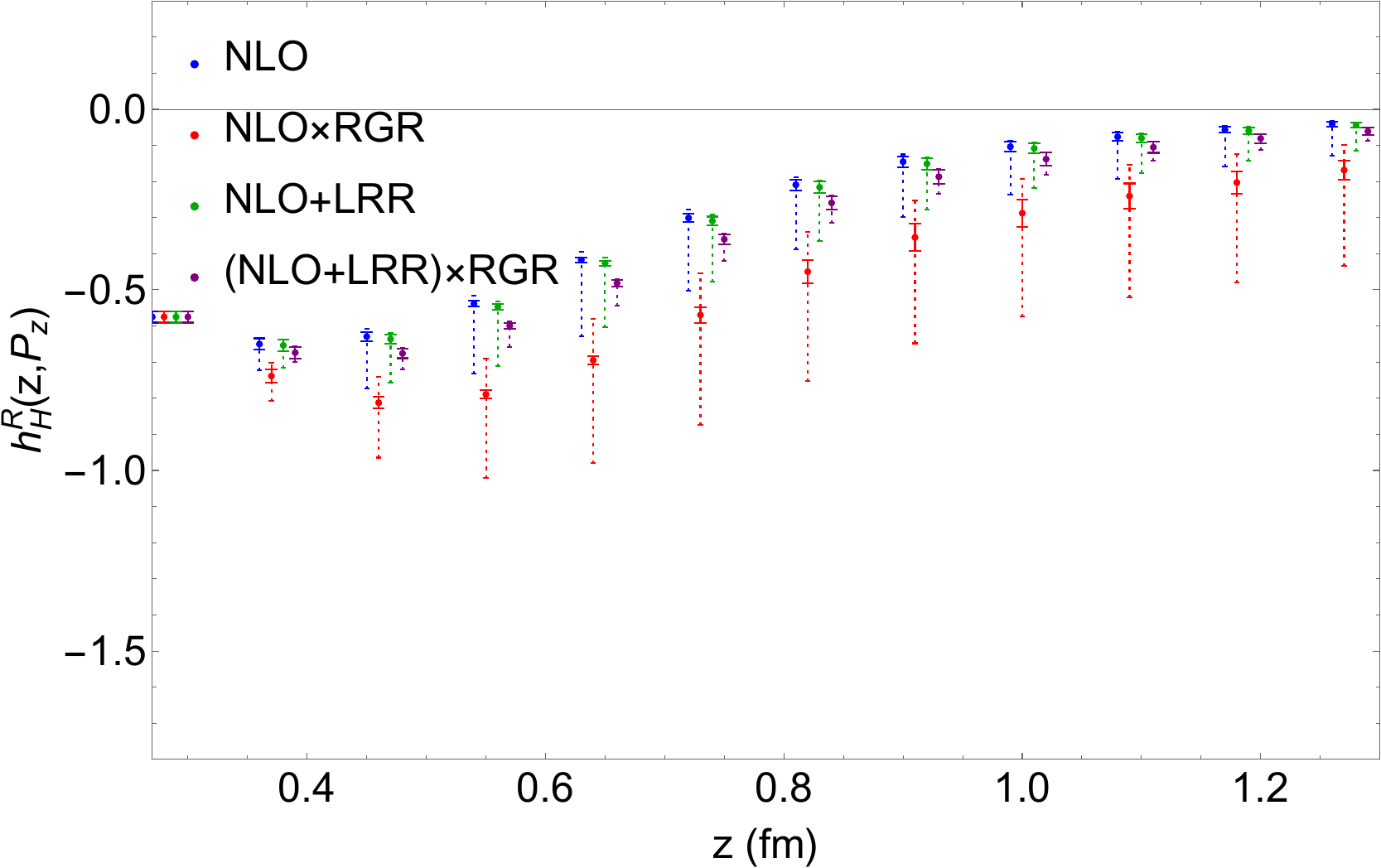}}
  \subfigure{\includegraphics[width=0.4\linewidth]{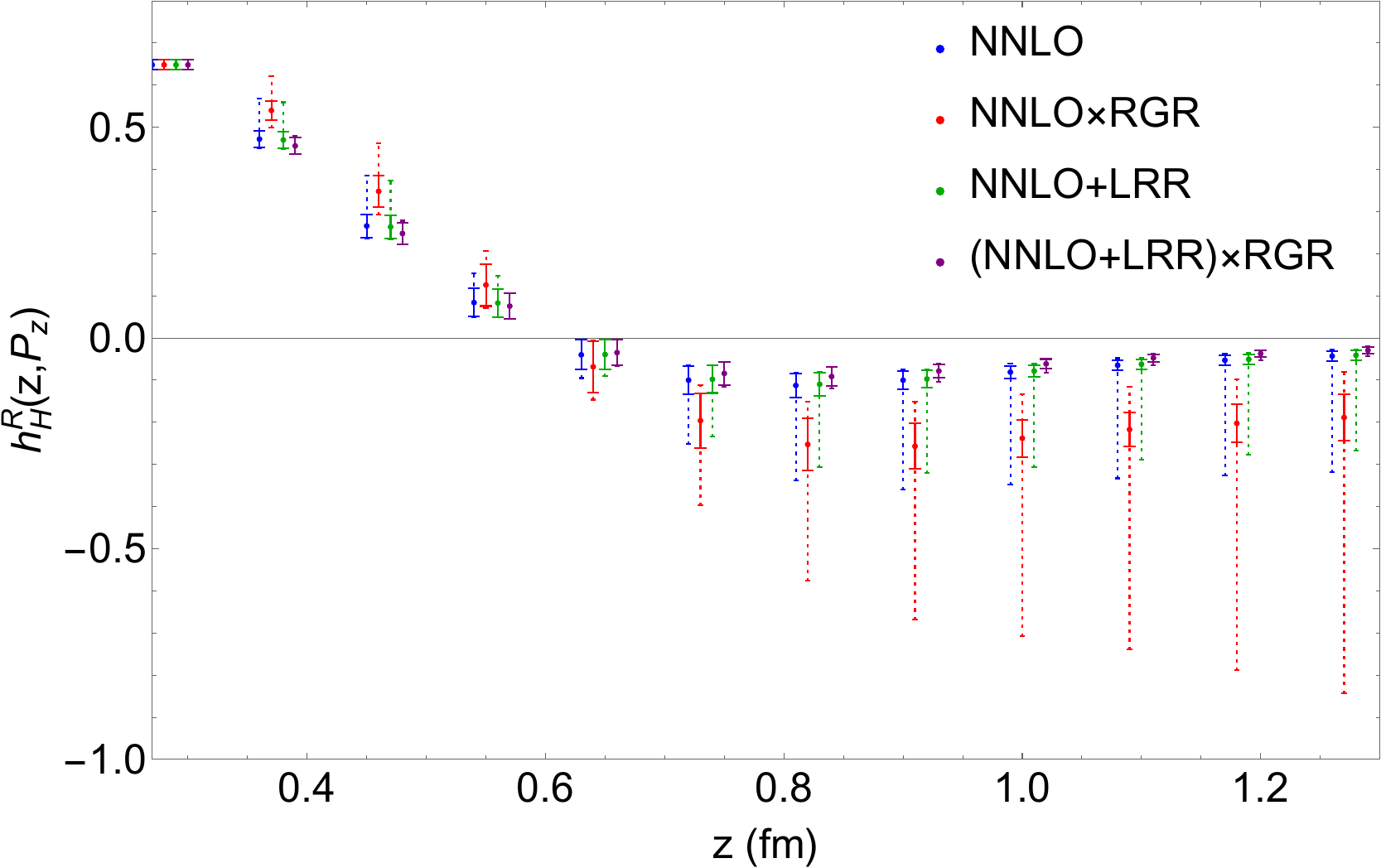}}\quad
  \subfigure{\includegraphics[width=0.4\linewidth]{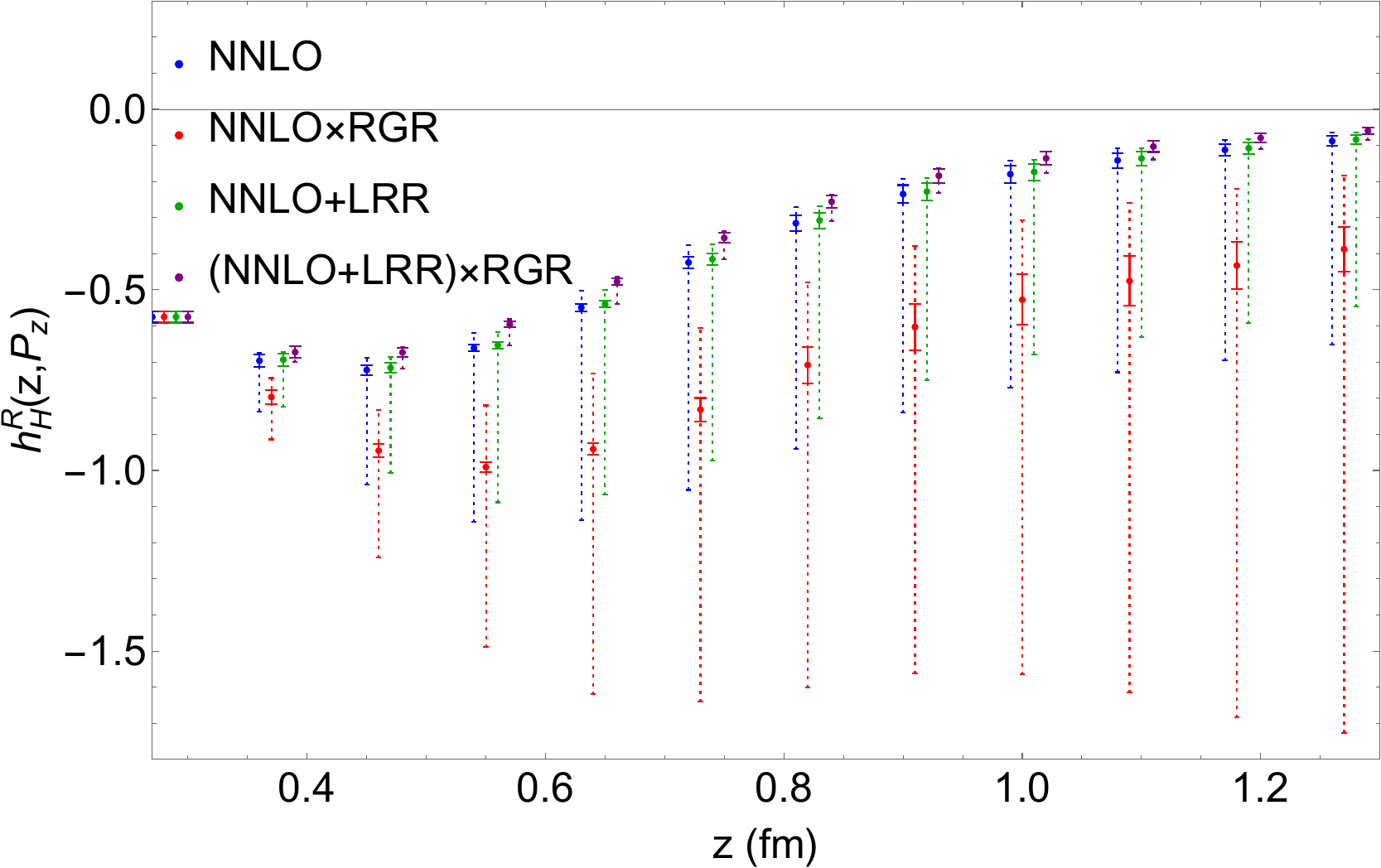}}
  \caption{Real (left column) and imaginary (right column) renormalized $h^R_H$ matrix elements at $Q^2=0$ with the
  top row showing data points of \N\ (blue), \NL\ (red), \NR\ (green) and \NLR\ (purple) improvements and bottom row with
  \NN\ (blue), \NNL\ (red), \NNR\ (green) and \NNLR\ (purple) improvements.
The solid error bars are statistical and the dashed error bars are combined statistical and systematic, the latter arising from the scale variation. Except for NLO and NNLO, the data points shown in the plots have been offset from their exact $z$ value to allow for readability.
  }\label{fig:xi0hRQ2-0}
\end{figure*}

We begin by looking into the special case of the nucleon unpolarized GPD at $Q^2=0$ and $\xi=0$, which is equivalent to the PDF.
We use the four different methods of computing the renormalon divergence $m_0$ at \N\ and again at \NN\ with the renormalized matrix elements $h^R_H(z,\xi=0,Q^2=0)$.
Our notation for the different schemes is ``(N)NLO$\times$RGR" for the RGR improvement only, ``(N)NLO+LRR" for the LRR improvement only, and ((N)NLO+LRR)$\times$RGR for both the RGR and LRR improvements.\footnote{Note that we adopt a different notation from Ref.~\cite{Zhang:2023bxs} to emphasize that the RGR process is applied to both the Wilson coefficient and the LRR modification as opposed to just the former.}
We show the real and imaginary parts of the matrix elements for (N)\N\ in the top (bottom) of Fig.~\ref{fig:xi0hRQ2-0}.
The (N)NLO, (N)NLO$\times$RGR, (N)NLO+LRR and ((N)NLO+LRR)$\times$RGR matrix elements are plotted in blue, red, green and purple, respectively.
Except for (N)NLO, the data points are offset slightly from their true $z$ values to allow for readability.
The plots contain both statistical error bars and combined statistical and systematic error bars from scale variation. 
In the case of the renormalized matrix elements, the systematic errors are computed by scale variation as was used in Ref.~\cite{Zhang:2023bxs}.
When RGR is applied to the Wilson coefficients, we vary the initial energy scale used in the RGR process, $c'\times\mathtt{z}^{-1}$, before evolving to the final desired one.
The central value corresponds to $c'=1.0$; the upper and lower error bars are derived by varying $c'$ from $0.75$ to $1.5$. The range $c'\in[0.75, 1.5]$ corresponds to a change of approximately 15\% on either side of $\alpha_s(\mu=2.0\text{ GeV})$.
This creates two additional curves with the maximum (minimum) value corresponding to the upper (lower) systematic error.
The systematic errors are asymmetric, since the strong-coupling dependence on energy scale is nonlinear.
When RGR is not applied to the Wilson coefficients, the systematic errors are determined by computing the renormalon divergence at energy scales 0.8~GeV and 2.8~GeV with 2.0~GeV being the central value.
These scale variations yield different measurements of the renormalon divergence and the upper- and lower-values are interpreted as the upper and lower systematic errors, respectively.
The RGR and LRR improvements to the Wilson coefficients give different central values and uncertainties in the renormalon divergence, resulting in different systematic errors in the renormalized matrix elements.

Examining the four \N\ schemes in the top row of Fig.~\ref{fig:xi0hRQ2-0}, we can see that the relative systematic errors are reduced by approximately 15\% to 35\% from \N\ to \NL.
The reduction from \NR\ to \NLR, however, is approximately 70\% to 90\% showing that leading-renormalon resummation has a much greater effect when used in combination with RGR.
This is to be expected, since the Wilson coefficients used to compute the renormalon divergence $m_0$ are series expansions in the strong coupling $\alpha_s$, and the renormalon divergence does not emerge until we expand the series to a power $n$ in the strong coupling where $n\sim 1/\alpha_s(\mu)$~\cite{Zichichi:1979gj,Beneke:1992ch}.
At our smallest energy scale used at fixed order, $\mu=0.8$~GeV, $\alpha_s(\mu)\approx 0.5$, and the renormalon divergence will not emerge unless we expand beyond quadratic terms in the strong coupling;
however, this does not mean that the renormalon divergence is irrelevant.
We can see that there is an increase up to fifteenfold in the absolute systematic errors from \N\ to \NR, since the latter does not account for the renormalon divergence.
When we compute the Wilson coefficients (and hence the renormalon divergence) at \NR, we set the initial energy scale to $\mu=\mathtt{z}^{-1}$.
At small $z$, this is a large energy scale, which results in a small $\alpha_s$, meaning that the renormalon divergence does not emerge at \NR\ in the series expansion.
The opposite occurs at large $z$ and, hence, the renormalon divergence can emerge at \NR.
This divergence is passed on to the calculation of the renormalon divergence, resulting in large systematic errors, particularly at large $z$, where the renormalon divergence occurs sooner in the series expansion.
For this reason, there is a significant difference between \N\ and \NR.
This reasoning also applies at \NN, in fact, to a greater extent, as can be seen in the bottom half of Fig.~\ref{fig:xi0hRQ2-0}.

With the above eight sets of renormalized matrix elements, we then construct the quasi-distributions.
First, we take each set of the real and imaginary renormalized matrix elements at large Wilson-line displacement and extrapolate them to infinite distance using Eq.~\ref{eq.Extrapolation}.
Here, we select the range $z\in[8a,15a]=[0.72, 1.35]$~fm for both the real and imaginary parts.
For all schemes as well as both real and imaginary parts, the $\chi^2/\text{dof}$ values are less than 1, which indicates the extrapolation formula is good model for unpolarized GPD matrix elements.
We construct a renormalized matrix element as a full function of $z$ by making a piecewise function.
At small $z$, we interpolate the lattice data, and at large $z$, we use the extrapolation model with the best-fit parameters.

We then Fourier transform our full function into momentum space using Eq.~\ref{eq.FourierTransform} to obtain the quasi-PDF and finally match to the lightcone using Eq.~\ref{eq.Matchingxi0}.
When RGR is not included in the calculation, the matching is performed at fixed order;
that is, $\mathcal{K}$ is evaluated at a fixed energy scale $\mu$.
When we include RGR, we perform matching at the energy scale $\mu=2xP_z$, which removes the large logarithms in the kernel, and then evolve to the desired scale with the DGLAP formula in Eq.~\ref{eq.DGLAP}.
The DGLAP equation begins to break down for $|x|\lesssim 0.2$, since the strong coupling $\alpha_s(\mu=2xP_z)$ becomes nonperturbative in this region.
Hence, we do not plot the unpolarized lightcone GPD data within this region and shade it in light gray.
In addition, the LaMET expansion breaks down for small- and large-$|x|$ as in the matching formula in Eq.~\ref{eq.Matchingxi0};
we approximate the region where these corrections become greater than or equal to one and shade these regions in dark gray.

The systematic errors for the unpolarized PDFs are computed by renormalizing the bare matrix elements with the upper and lower values of the renormalon divergence given by varying the scale.
We then perform the large-distance extrapolation, Fourier transformation and matching on the matrix elements.
When RGR matching is used, we set the initial scale to $\mu=c'\times 2xP_z$ with the central value corresponding to $c'=1.0$ and the systematic error bands coming from $c'=0.75$ and $1.5$ as in the determination of the renormalon divergence.
This gives us a central value for the PDF as well as two other values which correspond to the upper and lower systematic errors.

\begin{figure*}
    \centering
    \subfigure{\includegraphics[width=0.4\linewidth]{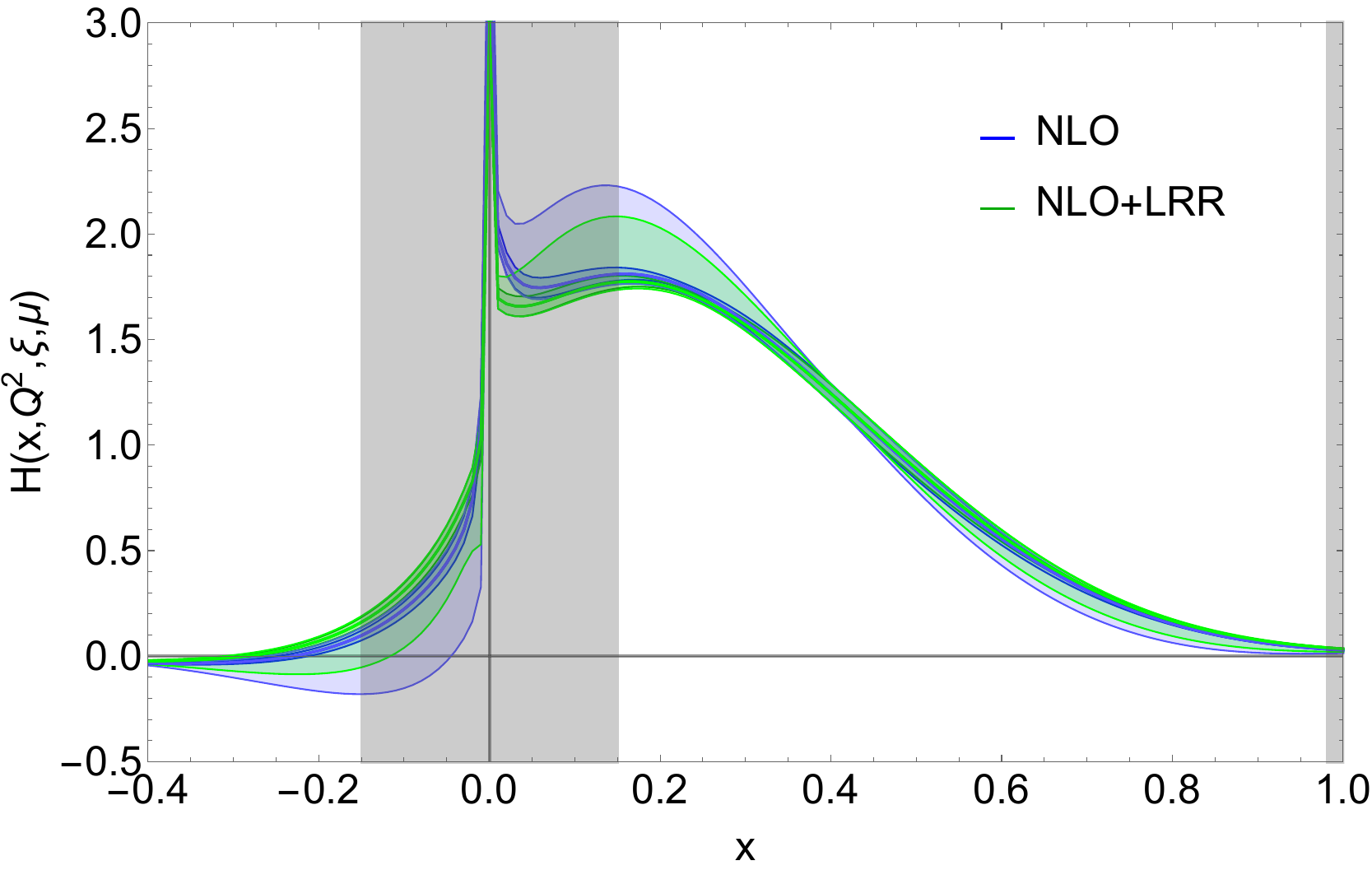}}\quad
    \subfigure{\includegraphics[width=0.4\linewidth]{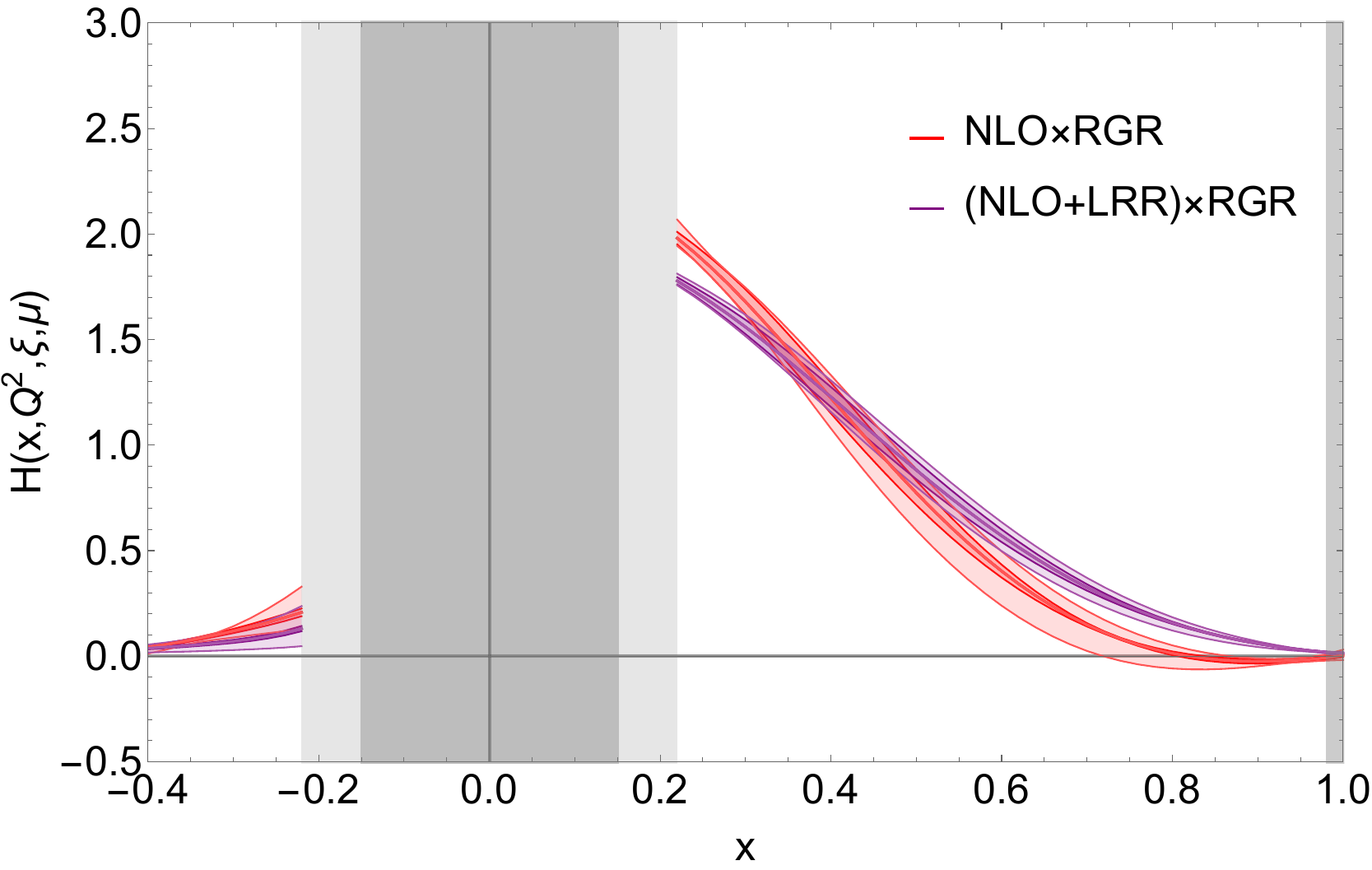}}
    \subfigure{\includegraphics[width=0.4\linewidth]{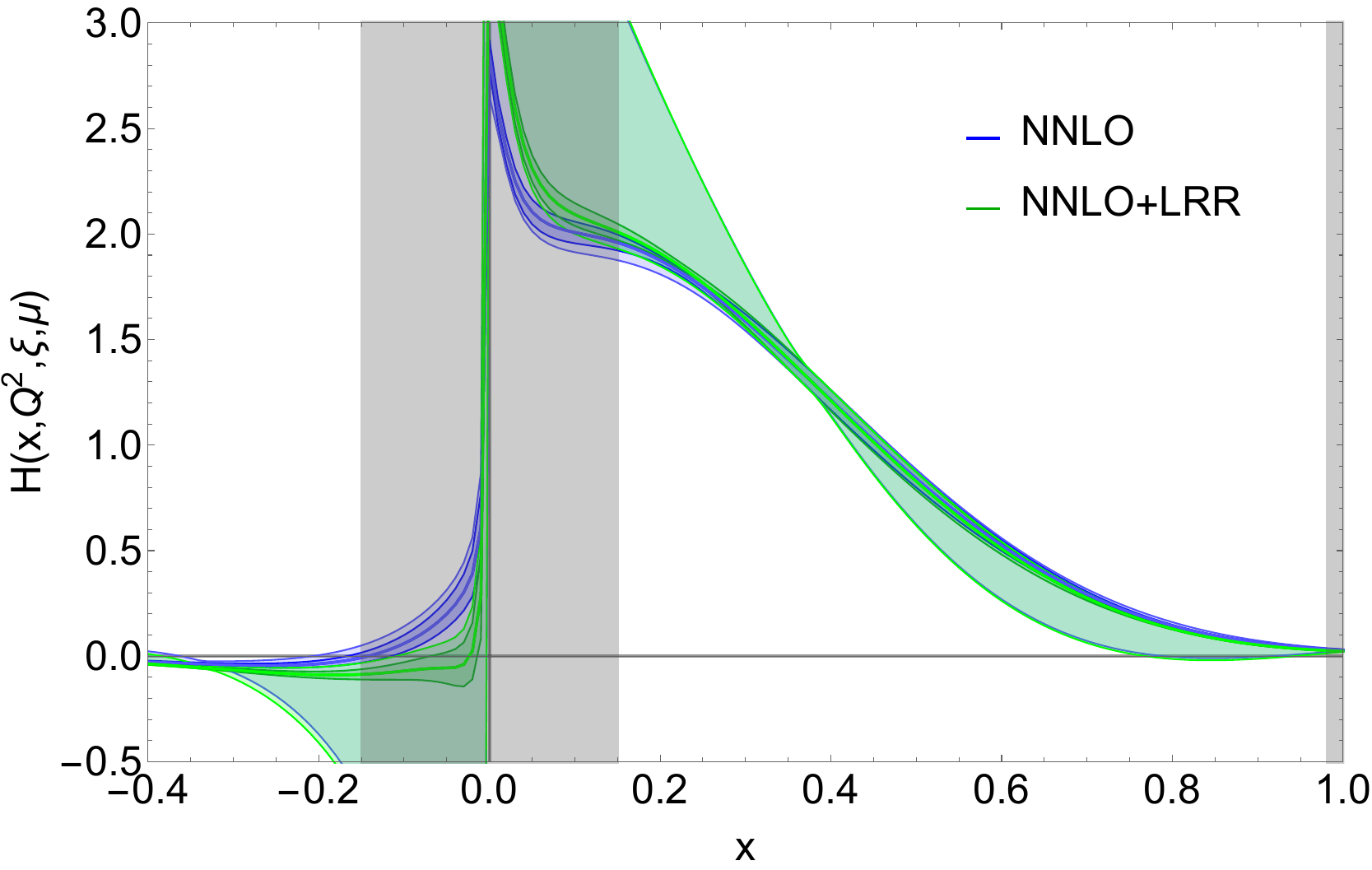}}\quad
    \subfigure{\includegraphics[width=0.4\linewidth]{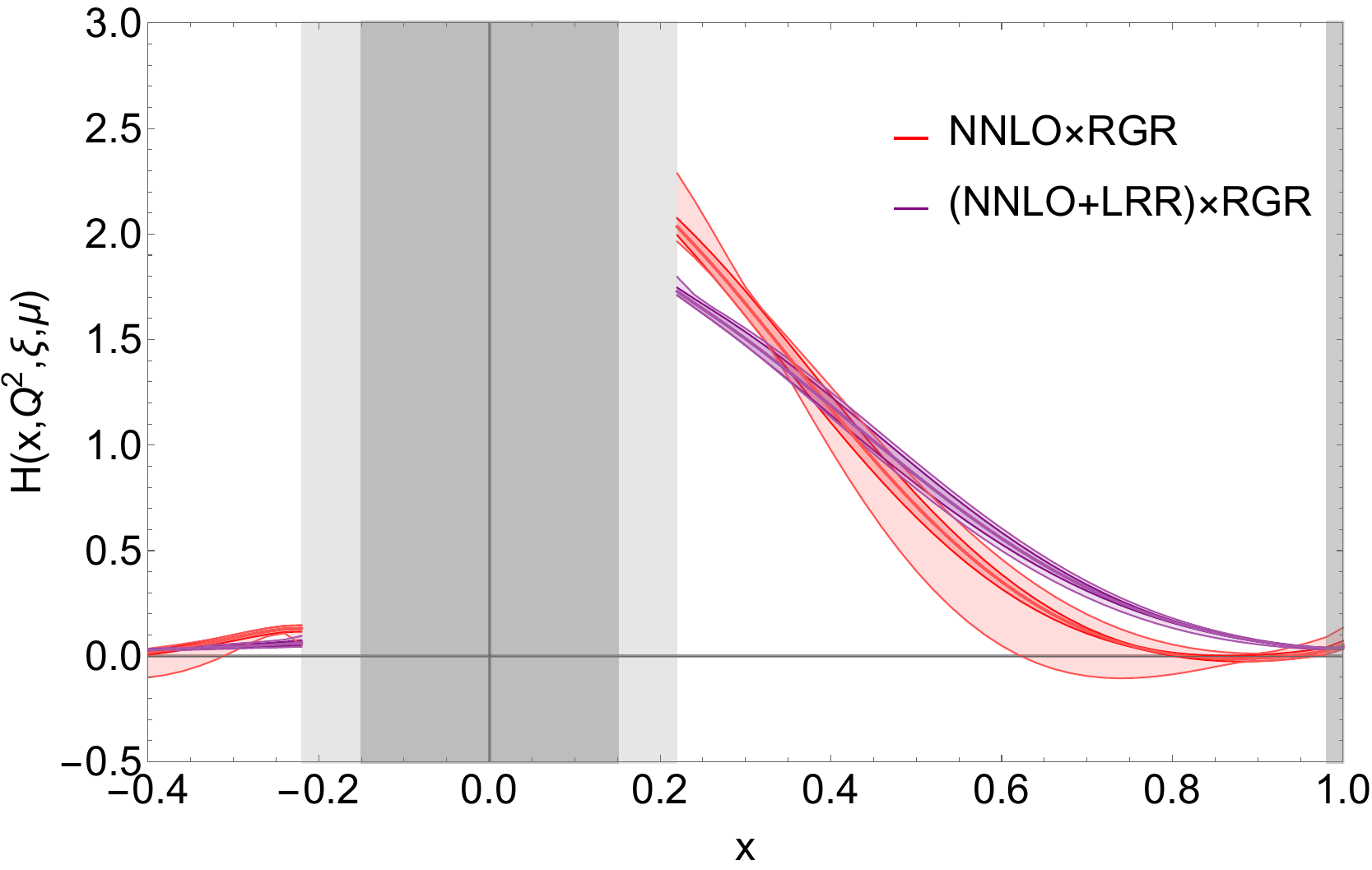}}\quad
    \caption{Isovector nucleon lightcone PDFs at NLO (top row) and NNLO (bottom row) without improvement (blue bands), with LRR only (green), with RGR only (red) and with both LRR and RGR (purple) improvements.
The dark-gray regions are the $x$ values at which the LaMET calculation breaks down.
In addition, when RGR is applied (right column), the matching formula breaks down for $|x|\lesssim 0.2$, so this region is shaded in light gray.}
    \label{fig:xi0HQ2-0}
\end{figure*}

In Fig.~\ref{fig:xi0HQ2-0} we show the lightcone unpolarized GPDs in the ``PDF limit" ($Q^2=0$ and $\xi=0$) with statistical errors (inner error bands) and combined statistical and systematic errors (outer error bands).
Since we have computed the unpolarized GPD, the regions $x>0$ and $x<0$ correspond to the combinations $F_u(x,Q^2,\xi)-F_d(x,Q^2,\xi)$ (``quark region") and $F_{\bar{d}}(x,Q^2,\xi)-F_{\bar{u}}(x,Q^2,\xi)$ (``antiquark region"), respectively.
The top (bottom) row shows the PDFs at (N)\N.
The left column shows no modifications and LRR only, the right column shows the RGR modification only and both RGR and LRR.
We plot the (N)NLO, (N)NLO$\times$RGR, (N)NLO+LRR and ((N)NLO+LRR)$\times$RGR PDFs in blue, red, green and purple, respectively.
In Eq.~\ref{eq.Matchingxi0}, there are corrections to the lightcone GPD that are suppressed with $P_z$ but grow at finite $P_z$ as $x\to 0$ or $|x|\to 1$.
We, therefore, shade in the regions at small and large $|x|$, where the LaMET calculation breaks down.

Examining the $x$-dependent GPDs, we consider first the four NLO schemes (top row of Fig.~\ref{fig:xi0HQ2-0}).
The statistical errors are more or less constant across the four of them, since the bare matrix elements are the same.
It is clear that the systematic errors are at their minimum when both the LRR and RGR improvements are applied simultaneously.
Indeed, much of the behavior of the systematic errors we see in Fig.~\ref{fig:xi0hRQ2-0} for the renormalized matrix elements also occurs in the PDFs.
Examining the large-$x$ region, we see that the four schemes become compatible with zero as $x\to 1$ within one to two sigma.
Across the quark region as a whole, we see that the central values across the four schemes are in general agreement for $x\gtrsim 0.3$ and the main difference is the variation in systematic errors.
We anticipate this result from the fact that the renormalized matrix elements differ in the renormalon divergence and its uncertainty;
the $m_0$ parameter in the four schemes are all compatible, but the error bars differ a great deal from one scheme to another.
The antiquark region, given the fluctuations across the different schemes, is compatible with zero.
Larger boost momenta will be required to improve the antiquark signal, as was demonstrated in Refs.~\cite{Chen:2018xof,Lin:2017ani,Lin:2018pvv,Liu:2018hxv}.
Hereafter, our focus will be on the quark region.

Turning next to the four NNLO results in the bottom half of Fig.~\ref{fig:xi0HQ2-0}, we see once again that the smallest systematic errors occur with \NNLR.
The systematic errors are approximately the same for \NN\ and \NNL, as we would expect from the renormalized matrix elements in Fig.~\ref{fig:xi0hRQ2-0} having similar systematic errors for the two schemes.
In the quark region, there is, in fact, little difference between the central values at \NLR\ and \NNLR\ except in the endpoint regions;
however, going to higher order reduces the systematic errors.
We note that while the \NNLR\ scheme has the smallest systematic errors, the central value remains consistent with both \NN\ and \NNL.
The central values of the \NNR\ results differ from the other three NNLO results due to the enhancement of the renormalon divergence when RGR is applied on its own.

Our results have shown that much of the advantage due to renormalization-group resummation and leading-renormalon resummation comes from a reduction in the systematic errors computed from scale variation.
The improved systematic errors with these schemes show that the benefits are transferable across different LaMET calculations, since their effects were first demonstrated in the case of the pion PDF~\cite{Zhang:2023bxs,Su:2022fiu} and pion DA~\cite{Holligan:2023rex}.
Given the significant differences in our (N)\NR\ and ((N)NLO+LRR)$\times$RGR~results and errors, we have shown that the renormalon divergence is a source of systematic errors that cannot be ignored as an esoteric phenomenon.

\subsection{Zero-Skewness $H$ and $E$ GPDs at $Q^2\neq 0$}

In this section, we examine our results for both the unpolarized zero-skewness $H$ and $E$ GPDs at nonzero momentum transfer.
The range of momentum transfer values used in this calculation is $Q^2\in\{0.19,0.39,0.77,0.97\}$~GeV$^2$.
We start this subsection by showing an example of the renormalized matrix elements at the intermediate value $Q^2=0.39$~GeV$^2$ to demonstrate the effects of LRR and RGR on the calculation.
The same procedures are applied to all of our $\xi=0$ GPD functions at all momentum transfers.
Since we have already studied the effects of \N, \NN\ and the applications of LRR and RGR in Sec.~\ref{subsec:Q0xi0}, we do not show every case here but restrict ourselves to \N, \NN, \NLR\ and \NNLR.

\begin{figure*}[htp]
  \centering
  \subfigure{\includegraphics[width=0.4\linewidth]{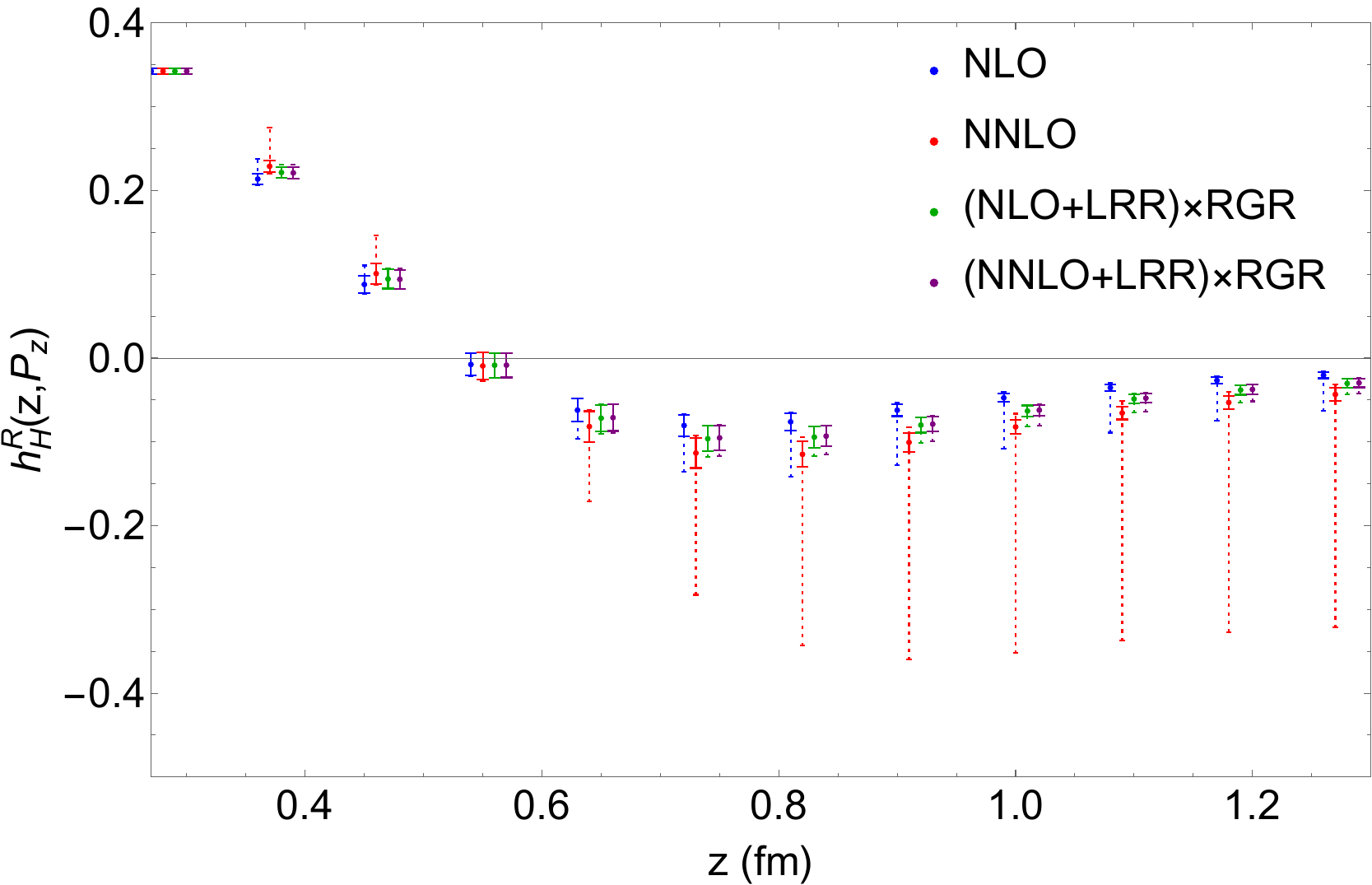}}\quad
  \subfigure{\includegraphics[width=0.4\linewidth]{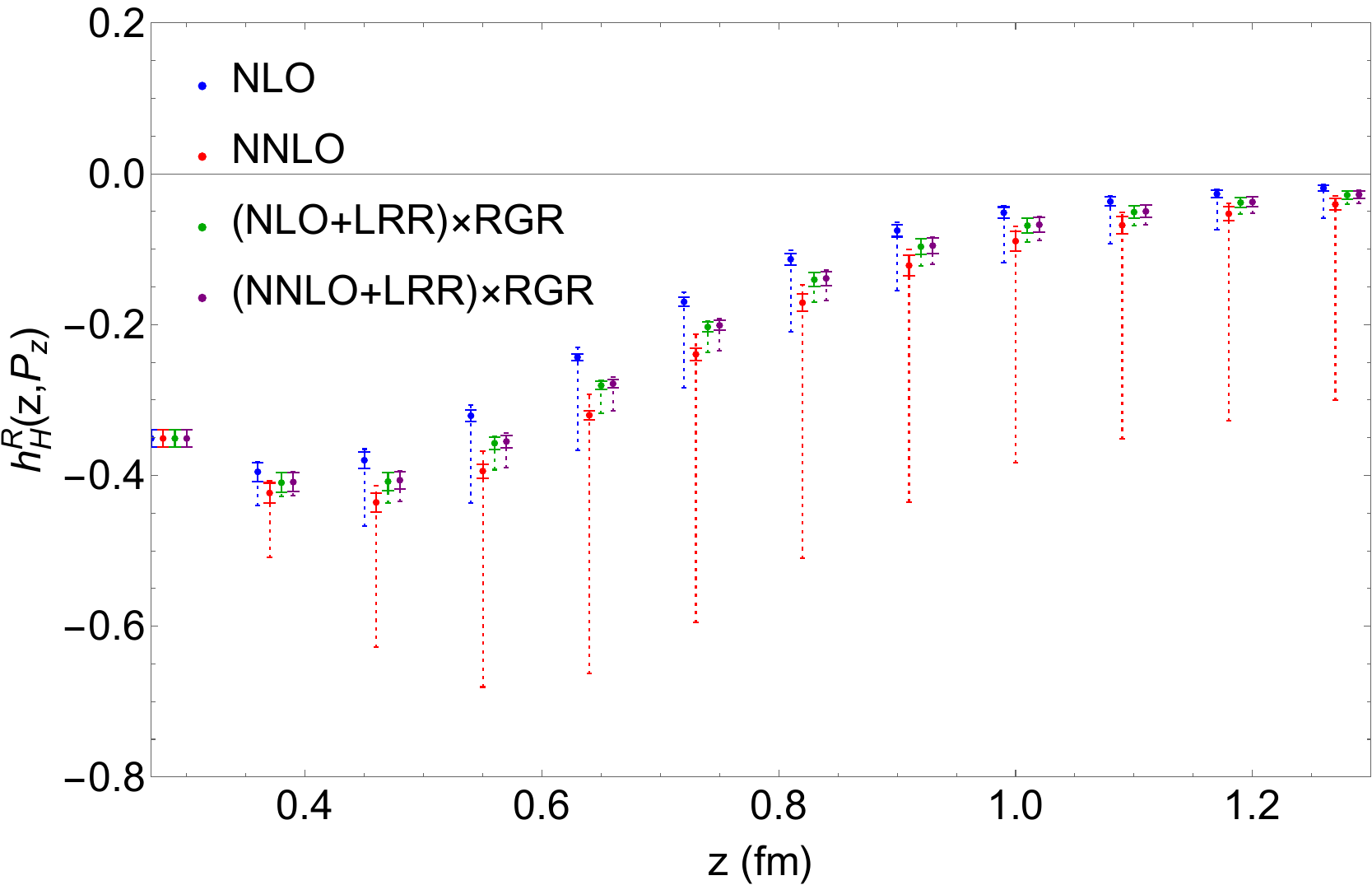}}
  \subfigure{\includegraphics[width=0.4\linewidth]{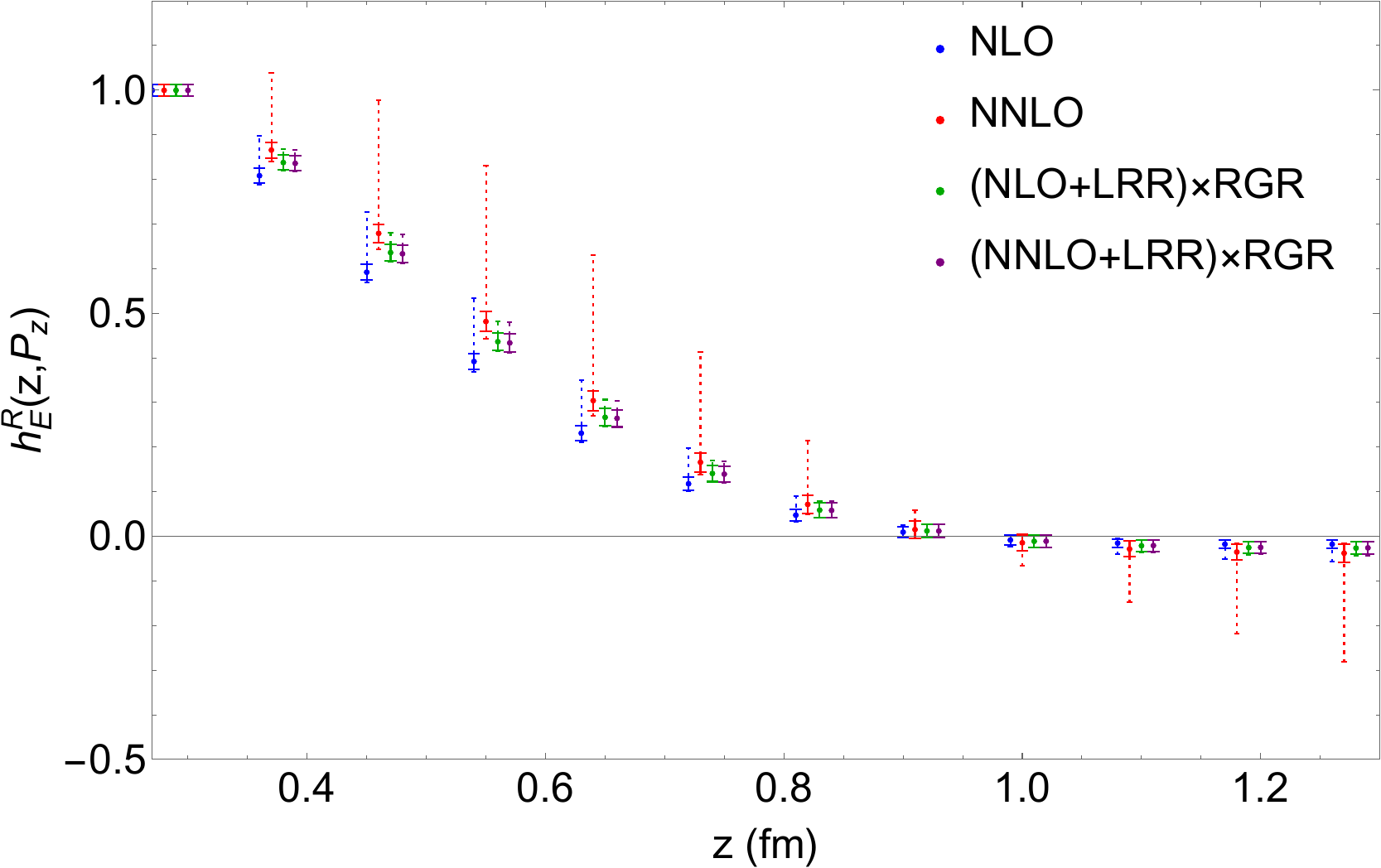}}\quad
  \subfigure{\includegraphics[width=0.4\linewidth]{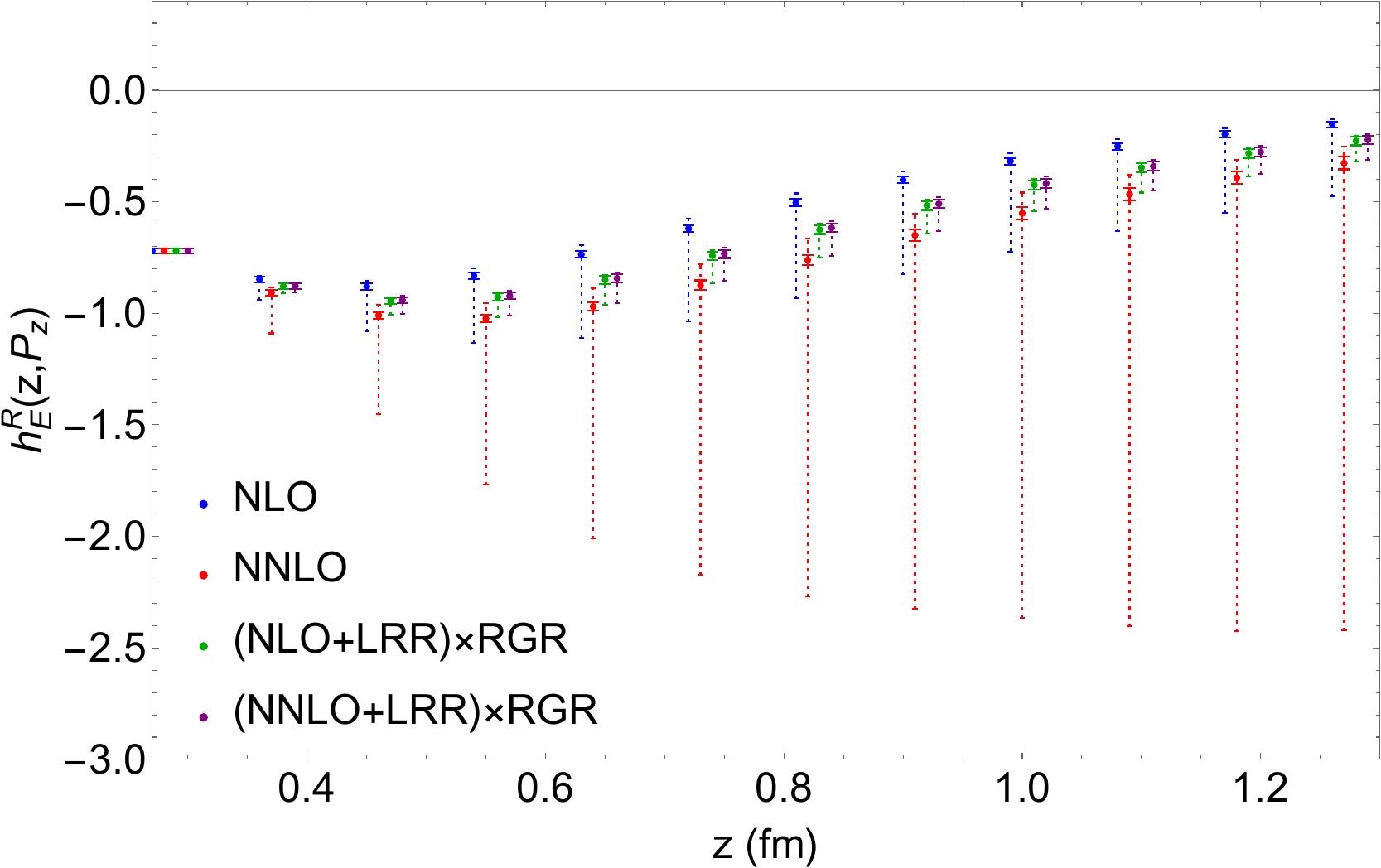}}
  \caption{Real (left column) and imaginary (right column) renormalized $h_H^R$ (top row) and $h_E^R$ (bottom row) matrix elements of \N\ (blue), \NN\ (red), \NLR\ (green) and \NNLR\ (purple) improvements at $Q^2=0.39$~GeV$^2$.
  The solid error bars are statistical and the dashed error bars are combined statistical and systematic, the latter arising from the scale variation.
  Except for \N\ (real and imaginary for both $h^R_H$ and $h^R_E$), the data points shown in the plots have been offset from their exact $z$ value to allow for readability.
}\label{fig:xi0hRQ2-0p39}
\end{figure*}

In Fig.~\ref{fig:xi0hRQ2-0p39}, we show the real (left column) and imaginary (right column) renormalized matrix elements for the $H$ and $E$ GPDs, $h_H^R$ (top row) and $h_E^R$ (bottom row), at zero skewness and $Q^2=0.39$~GeV$^2$.
At each $z$ point, two sets of errorbars are shown: the solid (inner) bars correspond to statistical errors and the dashed (outer) bars are statistical and systematic errors combined in quadrature.
The systematic errors are computed the same way as in Sec.~\ref{subsec:Q0xi0}.
Except for \N, the results are offset slightly from their true $z$ values to allow for readability.
Up to and including $Q^2=0.39$~GeV$^2$, we use the same fitting range for the large-distance extrapolation as was used in the PDF case ($Q^2=0$).
However, at $Q^2=0.77$ and 0.97~GeV$^2$, we use the fitting range $z\in[11a, 15a]=[0.99, 1.35]$~fm, since the $h^R_E$ matrix elements change sign at larger range for this momentum transfer, and such behaviour cannot be accommodated by the extrapolation model in Eq.~\ref{eq.Extrapolation}.

As in the case of the renormalized matrix elements at $Q^2=0$ in Fig.~\ref{fig:xi0hRQ2-0}, we see a significant decrease in systematic errors from \N\ to \NNLR\ (30\% to 70\%) and an even greater decrease from \NN\ to \NNLR\ (70\% to 90\%) in Fig.~\ref{fig:xi0hRQ2-0p39}.
This is to be expected, since the systematic errors of the renormalized matrix elements are governed by the renormalon divergence, which is itself determined by the Wilson coefficients.
The same benefits afforded by RGR and LRR that we see in Fig.~\ref{fig:xi0hRQ2-0} should occur at $Q^2=0.39$~GeV$^2$, since the same Wilson coefficients are used and improved in the same ways.
In addition, the systematic errors increase from NLO to NNLO as in the $Q^2=0$ case as we would expect from the behavior of the renormalon divergence.

\begin{figure*}[htp]
  \centering
  \subfigure{\includegraphics[width=0.4\linewidth]{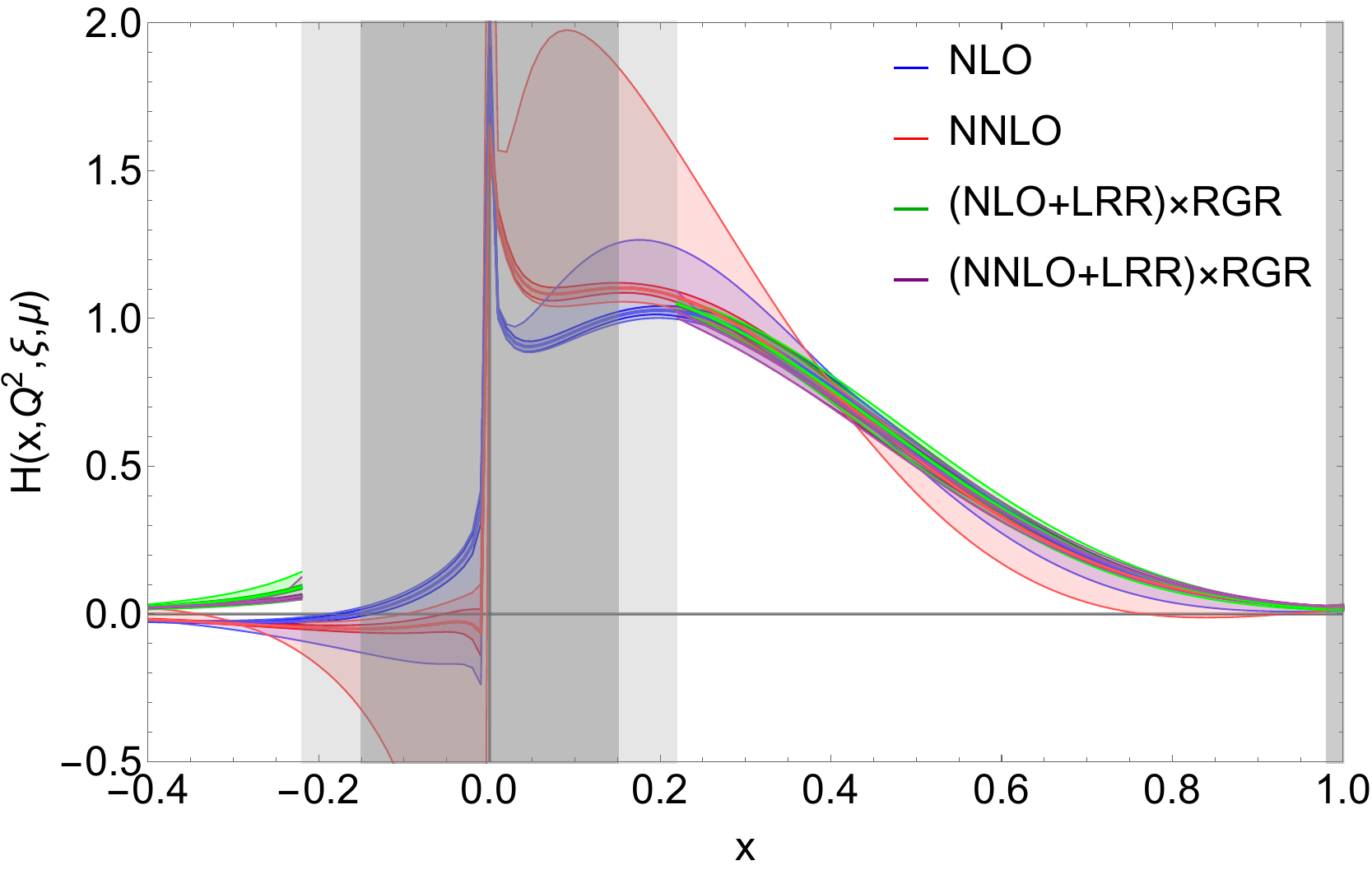}\quad
  \subfigure{\includegraphics[width=0.4\linewidth]{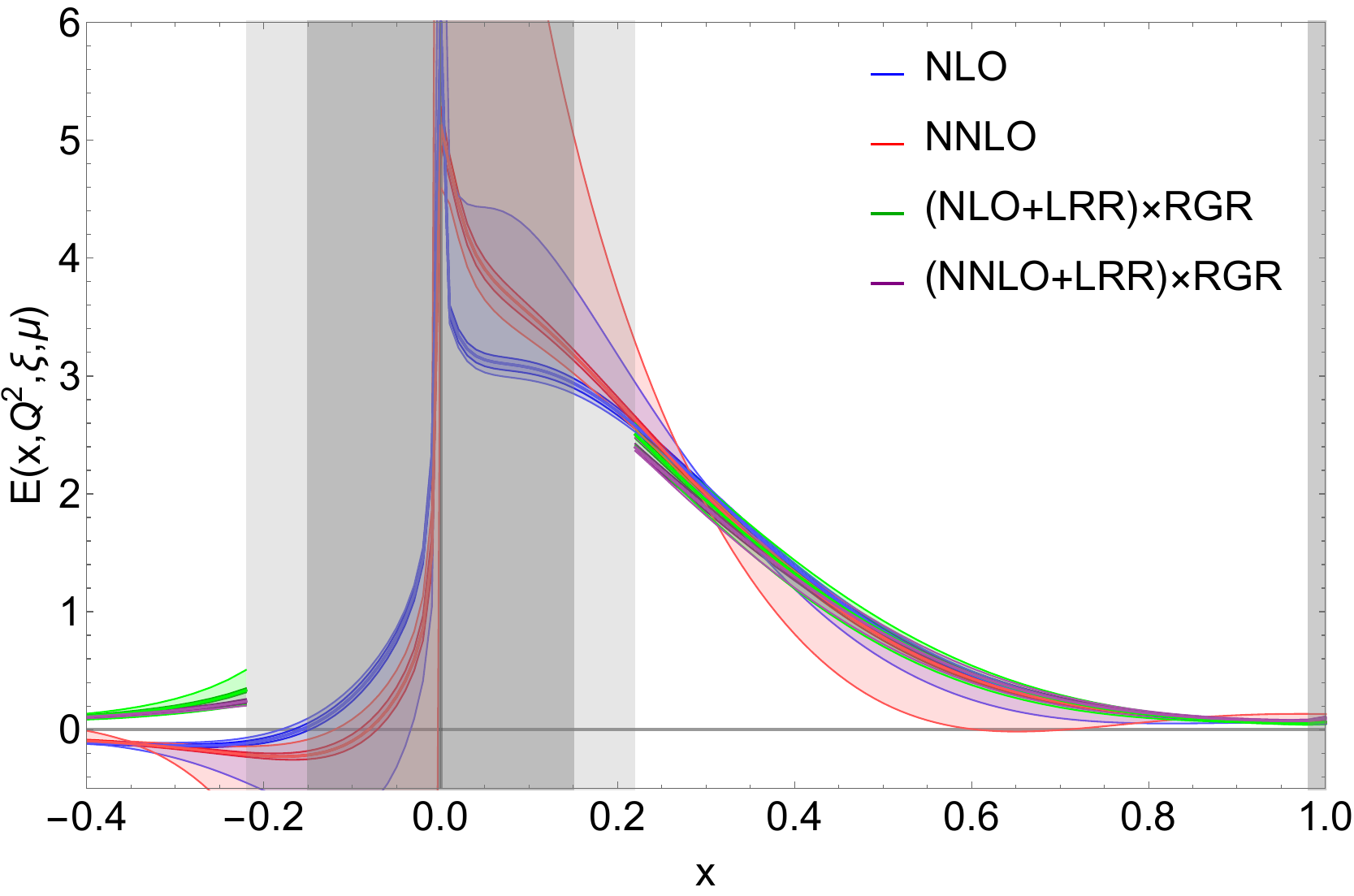}}}
  \caption{
   Lightcone $H$ and $E$ GPDs (left and right, respectively) with \N\ (blue), \NN\ (red), \NLR\ (green) and \NNLR\ (purple) evaluated at $Q^2=0.39$~GeV$^2$ and $\xi=0$.
   The inner bands are statistical errors;
   the outer bands are combined statistical and systematic errors, derived from the scale variation described in Sec.~\ref{subsec:Q0xi0}.
  The dark-gray regions are the $x$-values at which the LaMET calculation breaks down.
  In addition, when RGR is applied, the matching formula breaks down for $|x|\lesssim 0.2$, which is shaded in light gray.
  }\label{fig:xi0GPD-Q2-0p39}
\end{figure*}

In Fig.~\ref{fig:xi0GPD-Q2-0p39}, we show the unpolarized $H$ and $E$ GPDs in the \N, \NN, \NLR\ and \NNLR\ cases for $Q^2=0.39$~GeV$^2$.
The inner error bars are statistical and the outer error bars are combined statistical and systematic errors the latter computed in the same way as in the $Q^2=0$ case in Sec.~\ref{subsec:Q0xi0}.
As in the PDF case shown in Fig.~\ref{fig:xi0HQ2-0}, the systematics are at a minimum in the \NNLR\ scheme both for $H$ and $E$ GPDs.
The upper and lower systematic errors increase from \N\ to \NN\ for almost the whole interval $x\in[0.2, 0.8]$ which shows that the need to account for both the large logarithms and the renormalon divergence persists across different $Q^2$ values.
Also, the systematic errors decrease by up to 40\% from \NLR\ to \NNLR\ in the interval $x\in[0.3,0.9]$ both for $H$ and $E$ GPDs.
This shows the benefits of going up to two loops in the matching process.
Once again, the central values for all four schemes are in general agreement, showing that the main improvement afforded by RGR and LRR is a reduction in systematic errors.
It is also evidence for convergence in the matching procedure, since the central values for \NLR\ and \NNLR\ are close.
It is to be expected that the improved systematic errors persist across $Q^2$ values since the RGR and LRR improvements are universal and should be applicable in all LaMET calculations.
The fact that the systematics increase from \N\ to \NN\ and decrease from \NLR\ to \NNLR, shows again that the handling of systematic uncertainties must keep pace with higher orders in the matching and renormalization processes.

\begin{figure*}
    \centering
    \subfigure{\includegraphics[width=0.4\linewidth]{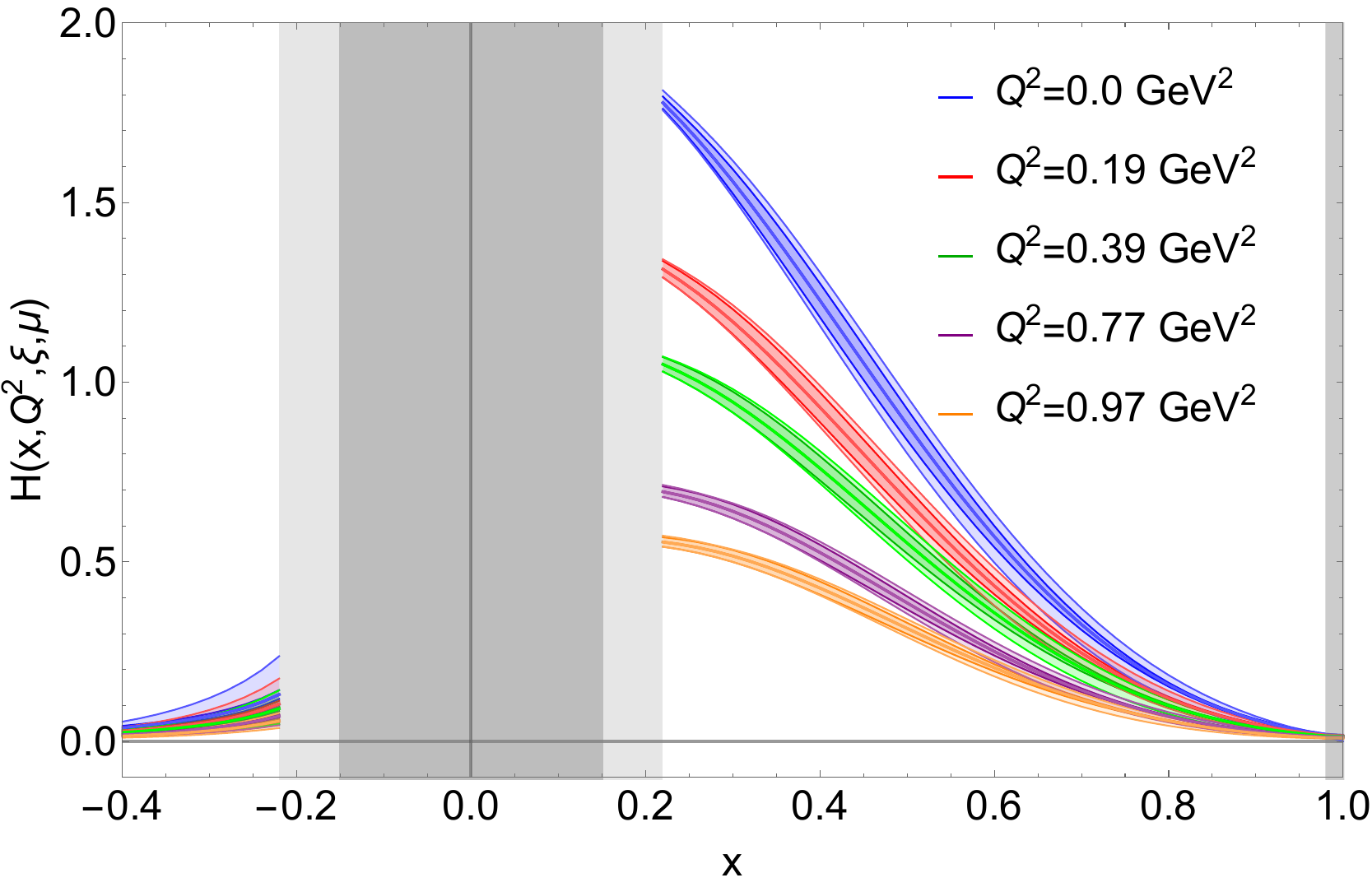}}\quad
    \subfigure{\includegraphics[width=0.4\linewidth]{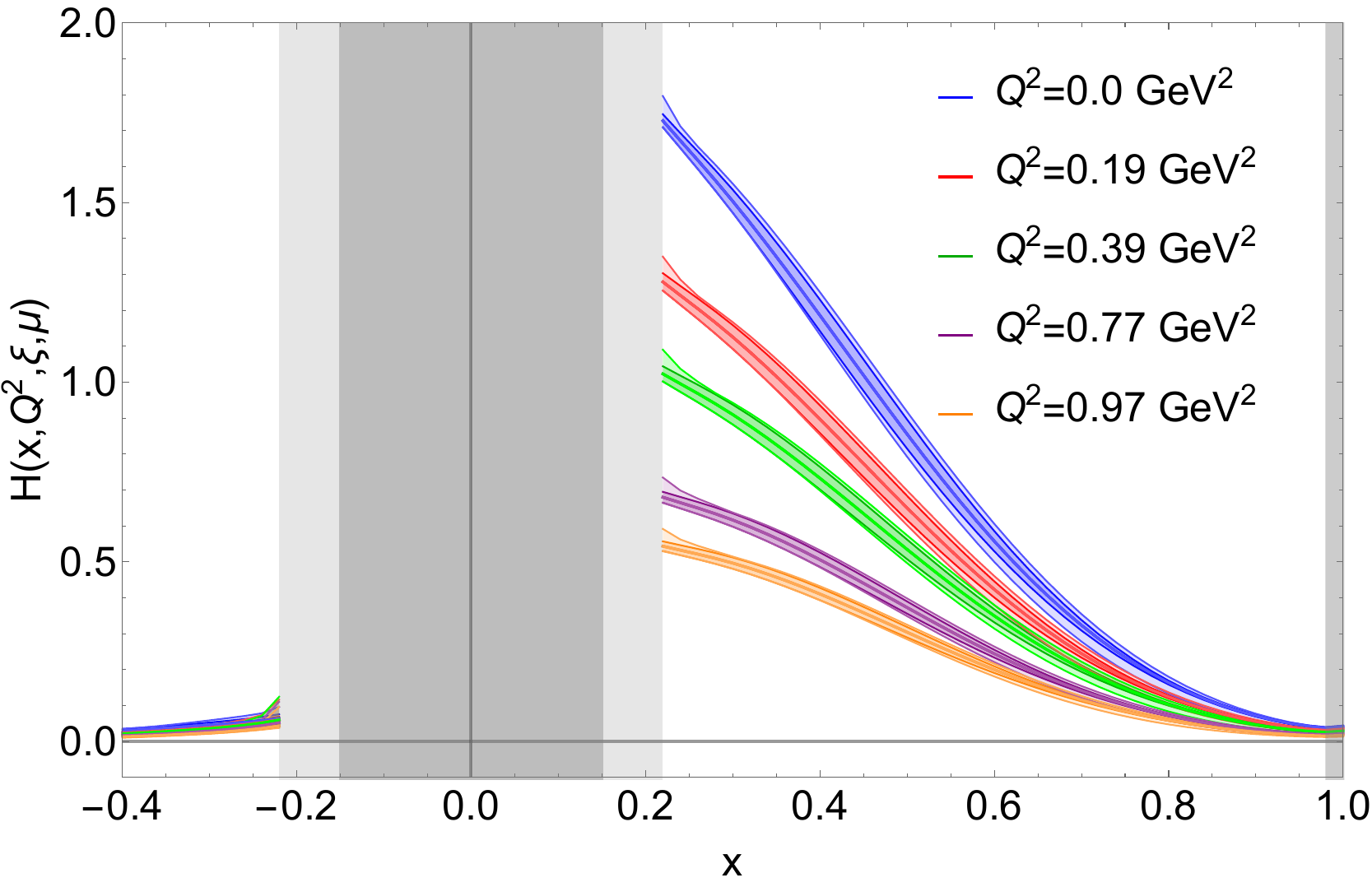}}
    \subfigure{\includegraphics[width=0.4\linewidth]{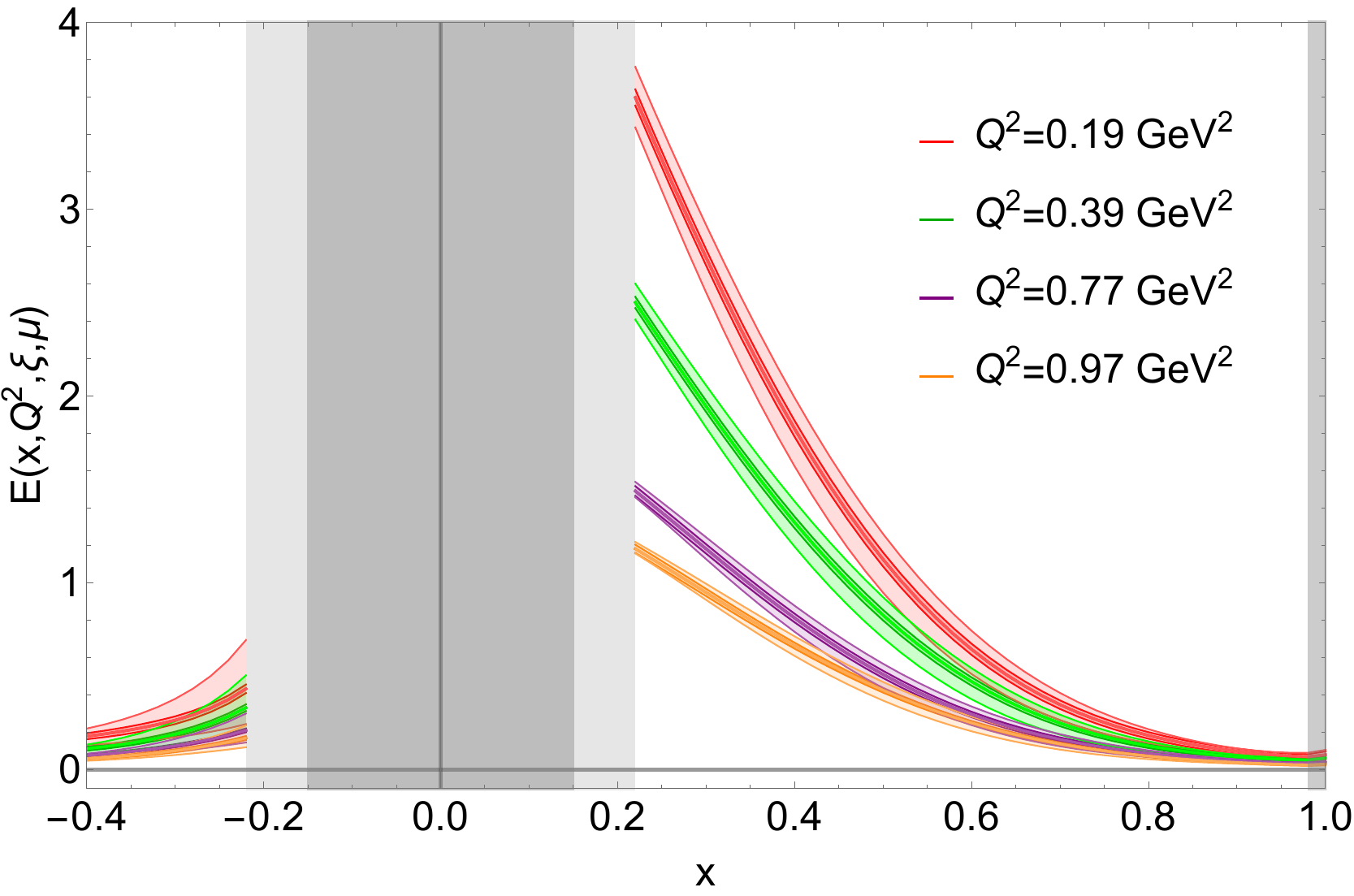}}\quad
    \subfigure{\includegraphics[width=0.4\linewidth]{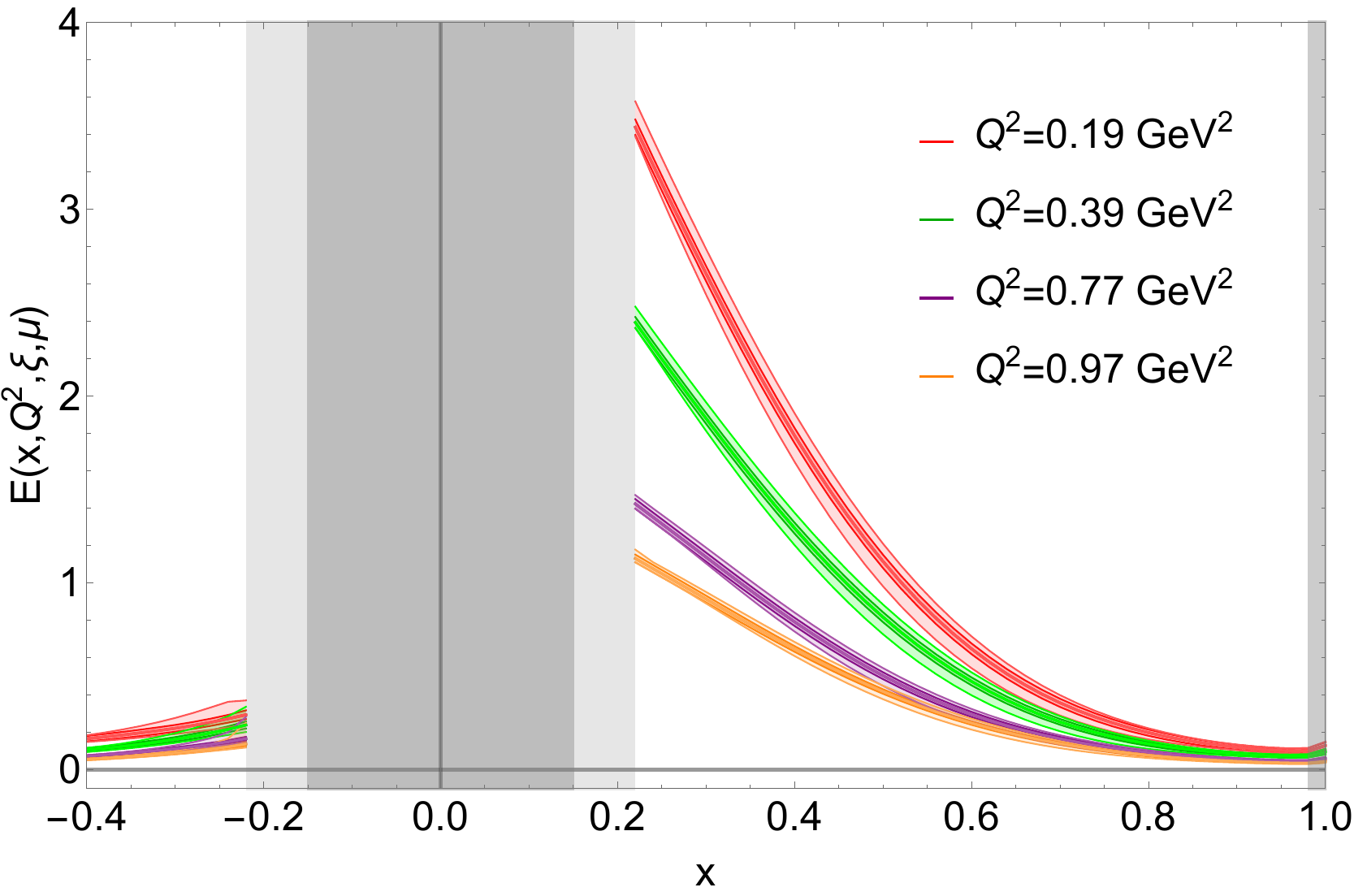}}
    \caption{\NLR\ (left column) and \NNLR\ (right column) $H$ (top row) and $E$ (bottom row) GPDs at $\xi=0$ and variable $Q^2$.
    The $Q^2\in\{0.0,0.19,0.39,0.77,0.97\}$~GeV$^2$ GPDs are plotted in blue, red, green, purple and orange, respectively.
    In all cases, the inner error bands are statistical and the outer error bands are combined statistical and systematic errors.
    The systematic errors decrease from \NLR\ to \NNLR, but in both cases are very small.
    The dark-gray regions are the $x$-values at which the LaMET calculation breaks down.
    In addition, when RGR is applied, the matching formula breaks down for $|x|\lesssim 0.2$, which is shaded in light gray.
    Note that the GPDs are suppressed as $Q^2$ increases.
    }
    \label{fig:xi0GPD-Q2-multiple}
\end{figure*}

These are the first applications of the RGR and LRR improvements to the LaMET calculation of the unpolarized nucleon GPD as well as the first application of hybrid-ratio renormalization to the same.
We plot both the $H$ and $E$ GPDs for $Q^2$ values from $0.19$ to $0.97$~GeV$^2$ (as well as the $H$ GPD for $Q^2=0$) at both \NLR\ and \NNLR\ in Fig.~\ref{fig:xi0GPD-Q2-multiple}.
The left (right) column corresponds to the unpolarized $H$ ($E$) GPD.
The top (bottom) row corresponds to ((N)NLO+LRR)$\times$RGR.
Once again, the inner bands correspond to statistical errors and the outer bands correspond to combined statistical and systematic errors computed as in Sec.~\ref{subsec:Q0xi0}.
We see that both the $H$ and $E$ GPDs decrease with $Q^2$, as has been seen in previous calculations of nucleon GPDs~\cite{Lin:2020rxa,Alexandrou:2020zbe}.
In all cases, the systematic errors are greatly reduced once again by the simultaneous additions of RGR and LRR.
This is more evidence for the universality of the renormalization-group resummation and leading-renormalon resummation.
The central values decrease from the quasi-GPD
to the \NLR\ GPD across all $Q^2$ and again when going from \NLR\ to \NNLR.
This is to be expected as the matching process tends to decrease the GPD value in the mid- to large-$x$ regions and increase the value at small-$x$. This is due to the probability of a parton carrying a high momentum-fraction decreasing as the hadron approaches the lightcone.
However, with the application of RGR in the matching, we cannot reliably study the small-$x$ region, $|x|\lesssim 0.2$.
Nevertheless, this first application of the RGR and LRR methods to GPDs at nonzero momentum transfer is a step toward precision GPDs from lattice QCD.

\section{Nonzero-Skewness GPDs}\label{sec:xinonzero}

In this section, we show the results for GPDs evaluated at nonzero skewness.
While the LRR method is directly transferable to $\xi\neq 0$, the RGR matching is not.
In $x$ space, the GPD is often broken down into two regions: the DGLAP region for $|x|>\xi$ and the ERBL region for $|x|<\xi$.
While the DGLAP evolution in Eq.~\ref{eq.DGLAP} is applicable to the corresponding region, a different scaling formula is required in the ERBL region, and there is the additional issue of two different intrinsic scales, which cannot be eliminated simultaneously by the judicious selection of a single initial energy.
For this reason, we only examine the effects of RGR on the renormalized matrix elements and confine our attention to the \N\ and \NL\ GPDs in momentum space.

\begin{figure*}[htp]
  \centering
  \subfigure{\includegraphics[width=0.4\linewidth]{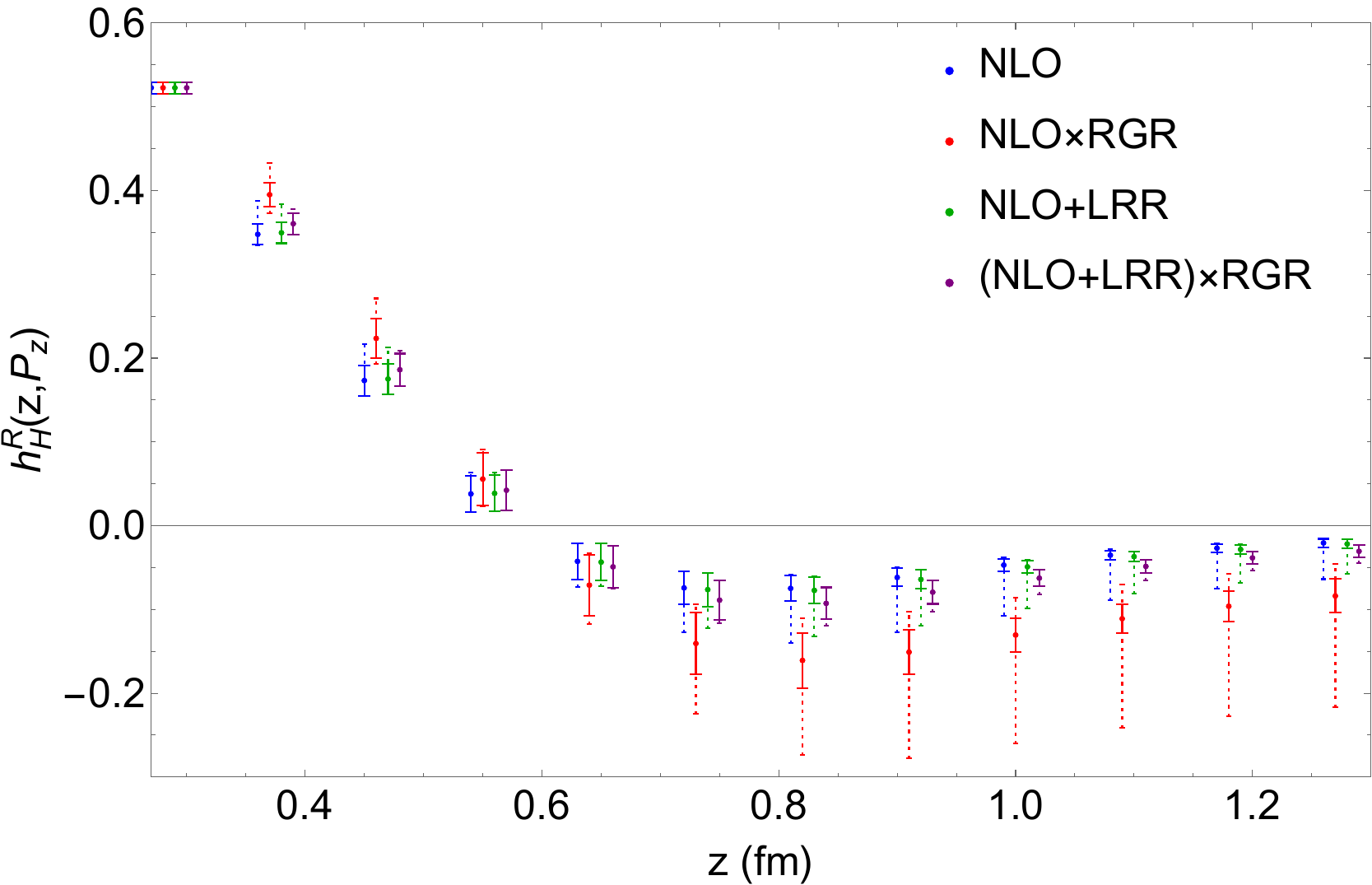}}\quad
  \subfigure{\includegraphics[width=0.4\linewidth]{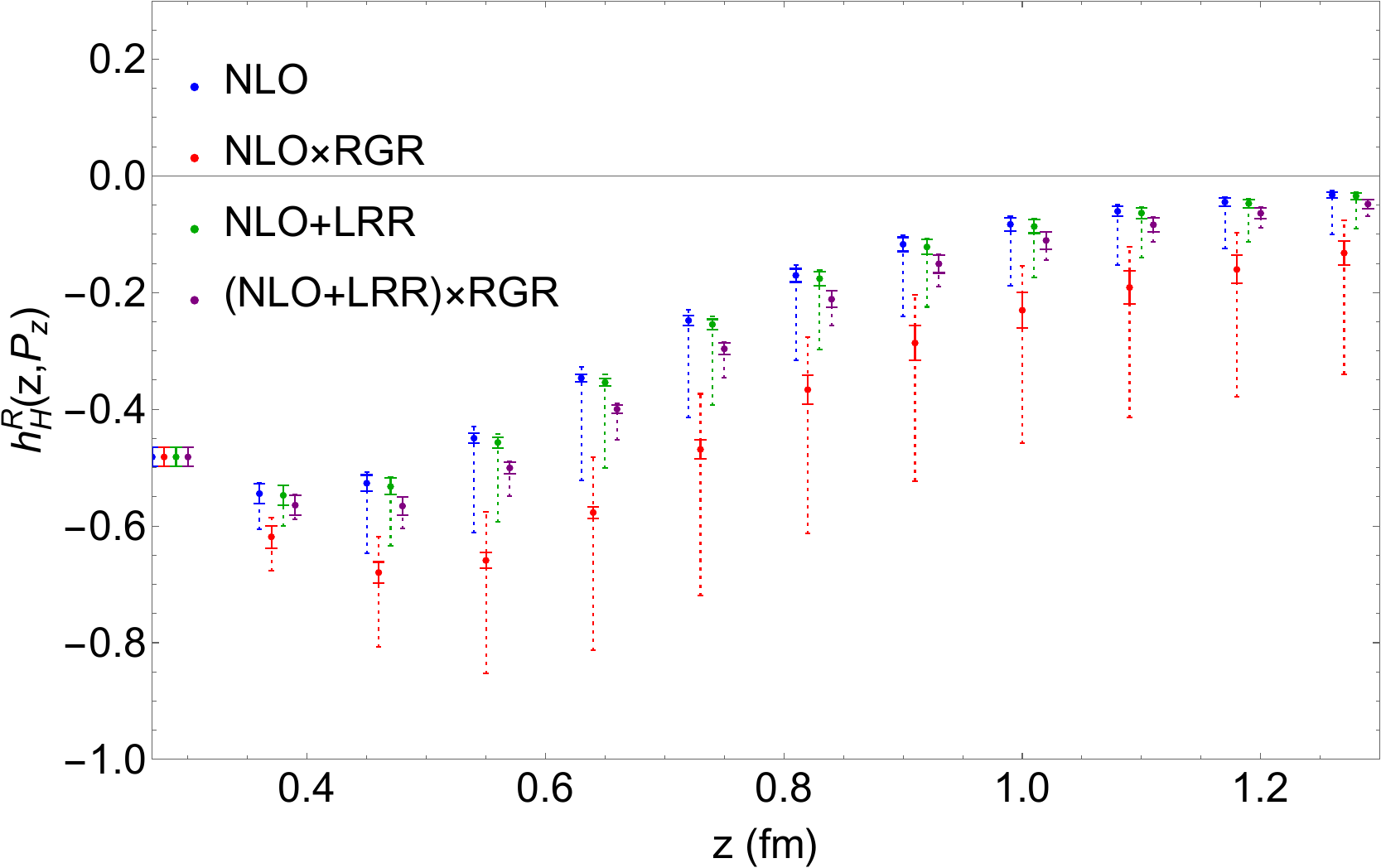}}
  \subfigure{\includegraphics[width=0.4\linewidth]{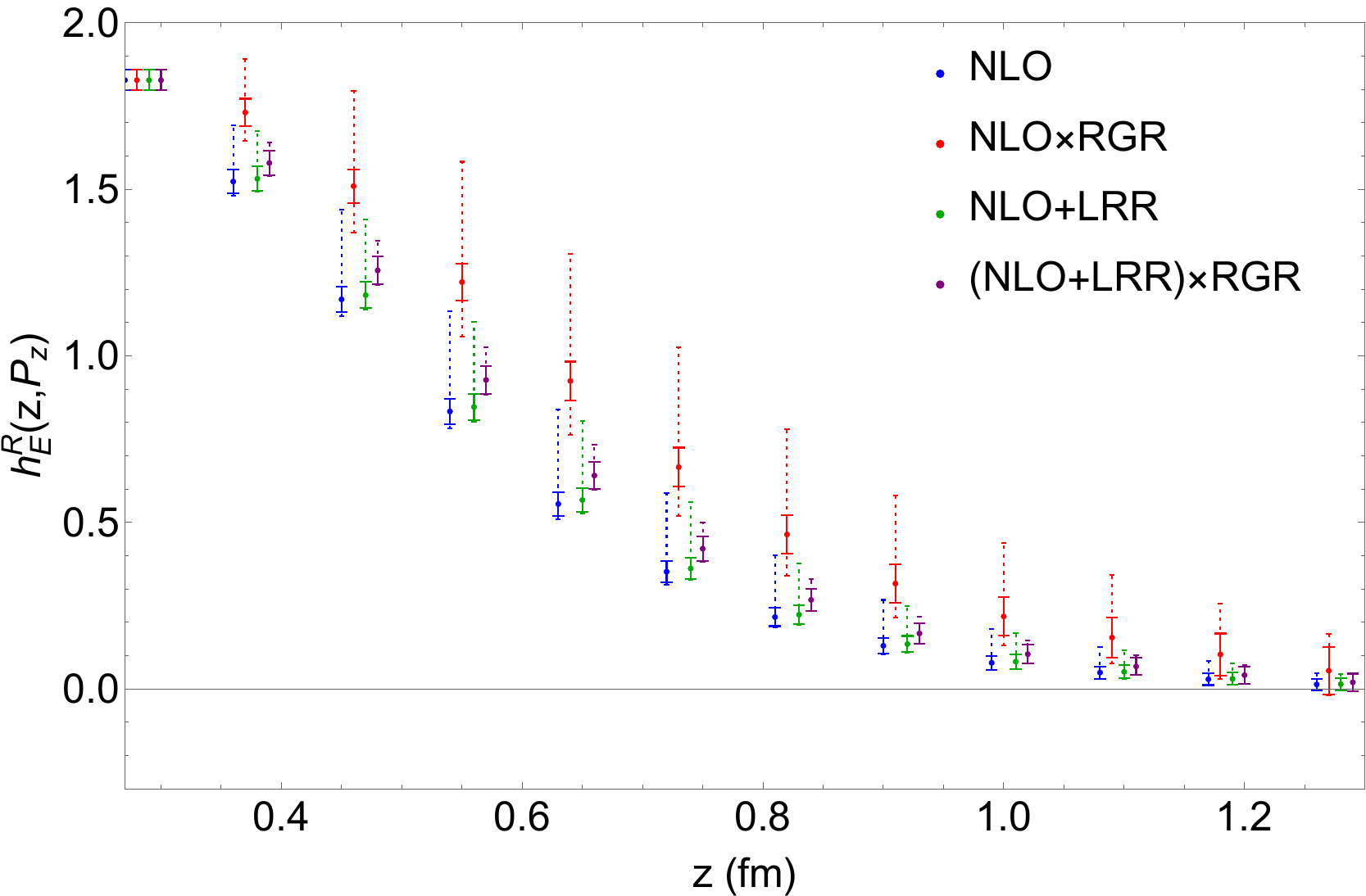}}\quad
  \subfigure{\includegraphics[width=0.4\linewidth]{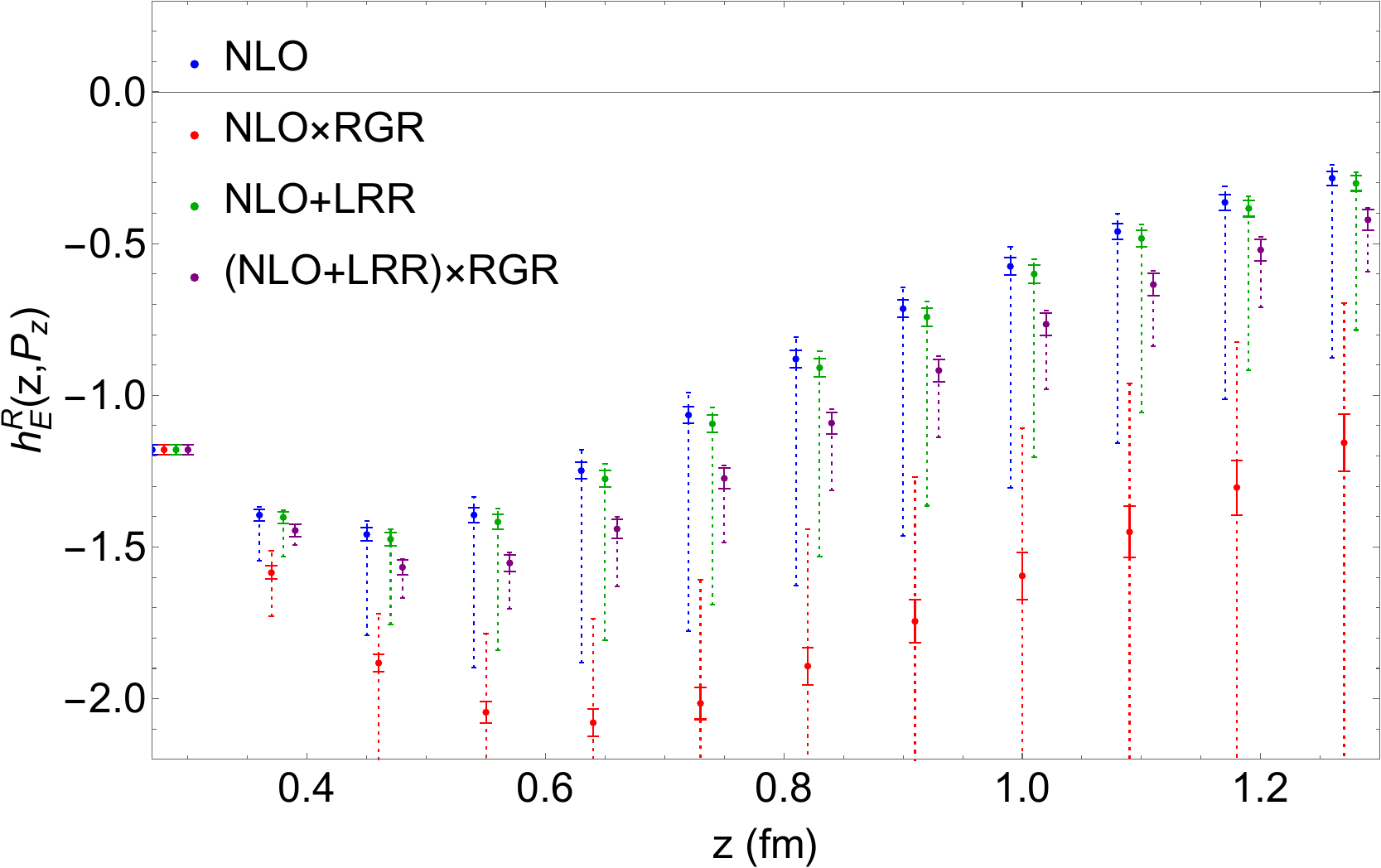}}
  \caption{
  Real (left column) and imaginary (right column) renormalized $h^R_H$ (top row) and $h^R_E$ (bottom row) matrix elements at $Q^2=0.23$ GeV$^2$ and $\xi=0.1$.
  We show data with \N\ (blue), \NL\ (red), \NR\ (green) and \NLR\ (purple) improvements.
The solid error bars are statistical and the dashed error bars are combined statistical and systematic, the latter arising from the scale variation.
Except for NLO (real and imaginary for both $h^R_H$ and $h^R_E$), the data points shown in the plots have been offset from their exact $z$ value to allow for readability.
}\label{fig:xi0p1hRQ2-0p23}
\end{figure*}

We start by looking at the renormalized matrix elements for both $h^R_H$ and $h^R_E$ at $Q^2=0.23$~GeV$^2$ and $\xi=0.1$ with statistical errors (inner bars) and combined statistical and systematic errors (outer bars) for all NLO four schemes in Fig.~\ref{fig:xi0p1hRQ2-0p23} with the outer systematic error bars computed as detailed in the previous sections.
The Wilson coefficients for the nonzero skewness are the same as those in zero skewness case.
Although we cannot yet apply RGR matching at nonzero skewness, we can see that the systematic errors in the renormalized matrix elements follow the same pattern as in the zero-skewness cases in Figs.~\ref{fig:xi0hRQ2-0} and \ref{fig:xi0hRQ2-0p39}.
This is evidence that the same improvements in the matching process adjusted for nonzero skewness should be equally effective as at $\xi=0$.

The matching kernel of the nonzero-skewness GPDs, $\mathcal{K}_\xi$, differs from the zero-skewness one used in the Eq.~\eqref{eq.Matchingxi0} due to the fact that skewness parameter $\xi$ encapsulates the change in the struck hadron's longitudinal momentum.
To date, the $\xi\neq 0$ matching kernel,  $\mathcal{K}_\xi$, has only been computed up to NLO for unpolarized GPDs in the hybrid-ratio scheme in Refs.~\cite{Ji:2015qla,Xiong:2015nua,Liu:2019urm,Yao:2022vtp}; however,  the kinematic setup of the kernels in Refs.~\cite{Ji:2015qla,Xiong:2015nua,Liu:2019urm} in the ERBL region are incomplete.
For this work, we adopt the $\xi\neq 0$ matching kernel, $\mathcal{K}_\xi$, from Ref.~\cite{Yao:2022vtp}:
\begin{widetext}
\begin{multline}\label{eq.xinonzero}
 \frac{1}{|y|}\mathcal{K}_\xi(x,y,\mu,\xi,P_z)=\delta(x-y)+\frac{\alpha_s(\mu)C_F}{4\pi}\left[\left(\frac{|\xi+x|}{2\xi(\xi+y)}+\frac{|\xi+x|}{(\xi+y)(y-x)}\right)\left(\ln\left(\frac{4(\xi+x)^2P_z^2}{\mu^2}\right)-1\right)\right.\\
	+\left.\left(\frac{|\xi-x|}{2\xi(\xi-y)}+\frac{|\xi-x|}{(\xi-y)(x-y)}\right)\left(\ln\left(\frac{4(\xi-x)^2P_z^2}{\mu^2}\right)-1\right)\right.\\
	\left.+\left(\left(\frac{\xi+x}{\xi+y}+\frac{\xi-x}{\xi-y}\right)\frac{1}{|x-y|}-\frac{|x-y|}{\xi^2-y^2}\right)\left(\ln\left(\frac{4(x-y)^2P_z^2}{\mu^2}\right)-1\right)\right].
\end{multline}
\end{widetext}
Note that we modified the kernel to convention in which there is an extra factor of $1/|y|$ in the integrand, whereas Ref.~\cite{Yao:2022vtp} absorbed this factor into the kernel itself.
The nonzero skewness matching kernel $\mathcal{K}_\xi$ contains singularities at $y=0$, $y=x$ and $|y|=\xi$.
It is invariant under $\xi\to -\xi$ and recovers the NLO zero-skewness kernel when taking limit $\xi\to 0$.
The LRR modification to the matching kernel is the same at both zero and nonzero skewness~\cite{Zhang:2023bxs}.
The LRR matching modification is derived from the LRR modification to the Wilson coefficients.
Since the Wilson coefficients are the same for zero and nonzero skewness, the same modification to the matching kernel is applicable.

\begin{figure*}[htp]
\centering
\subfigure{\includegraphics[width=0.4\linewidth]{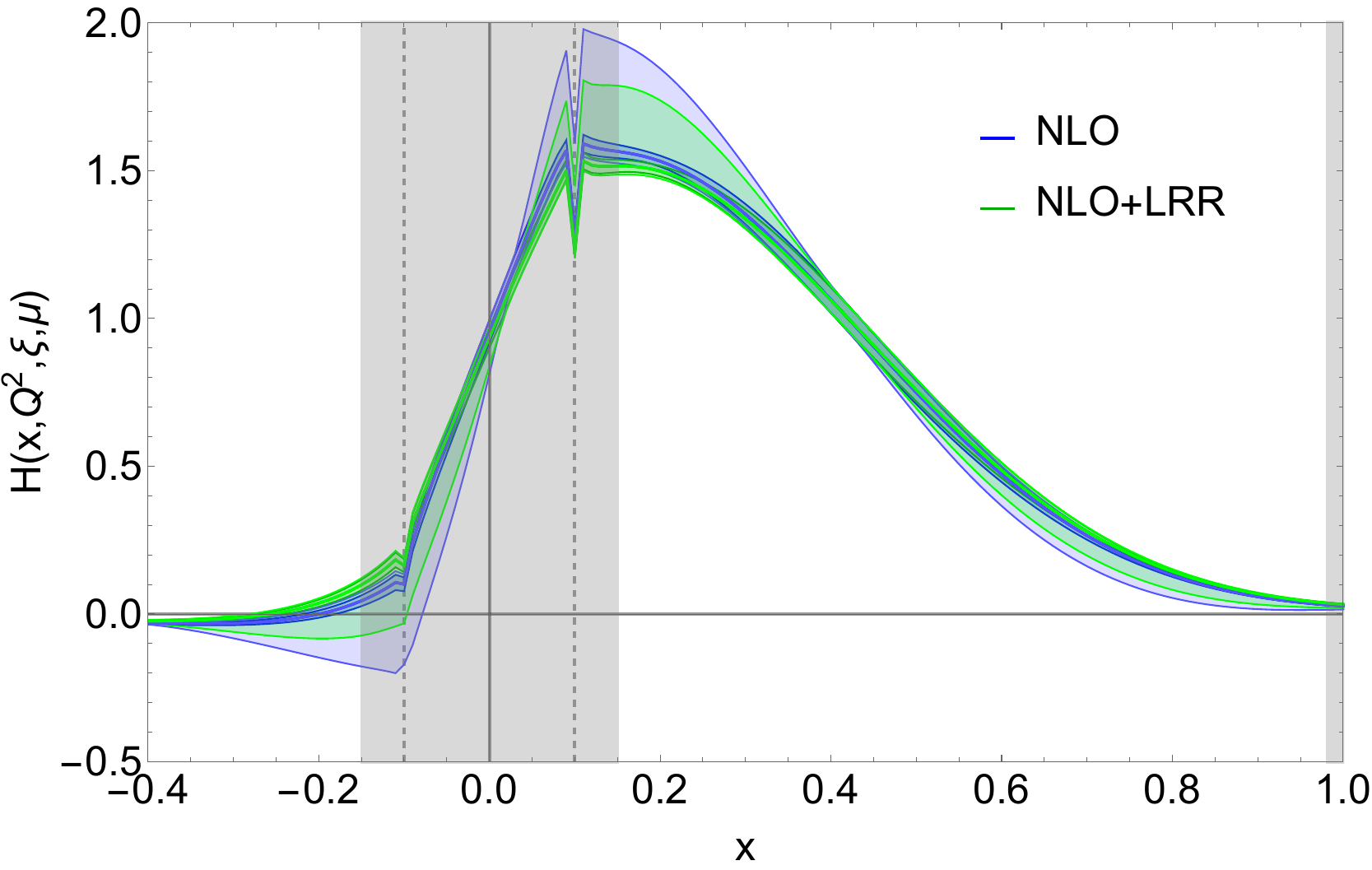}}\quad
\subfigure{\includegraphics[width=0.4\linewidth]{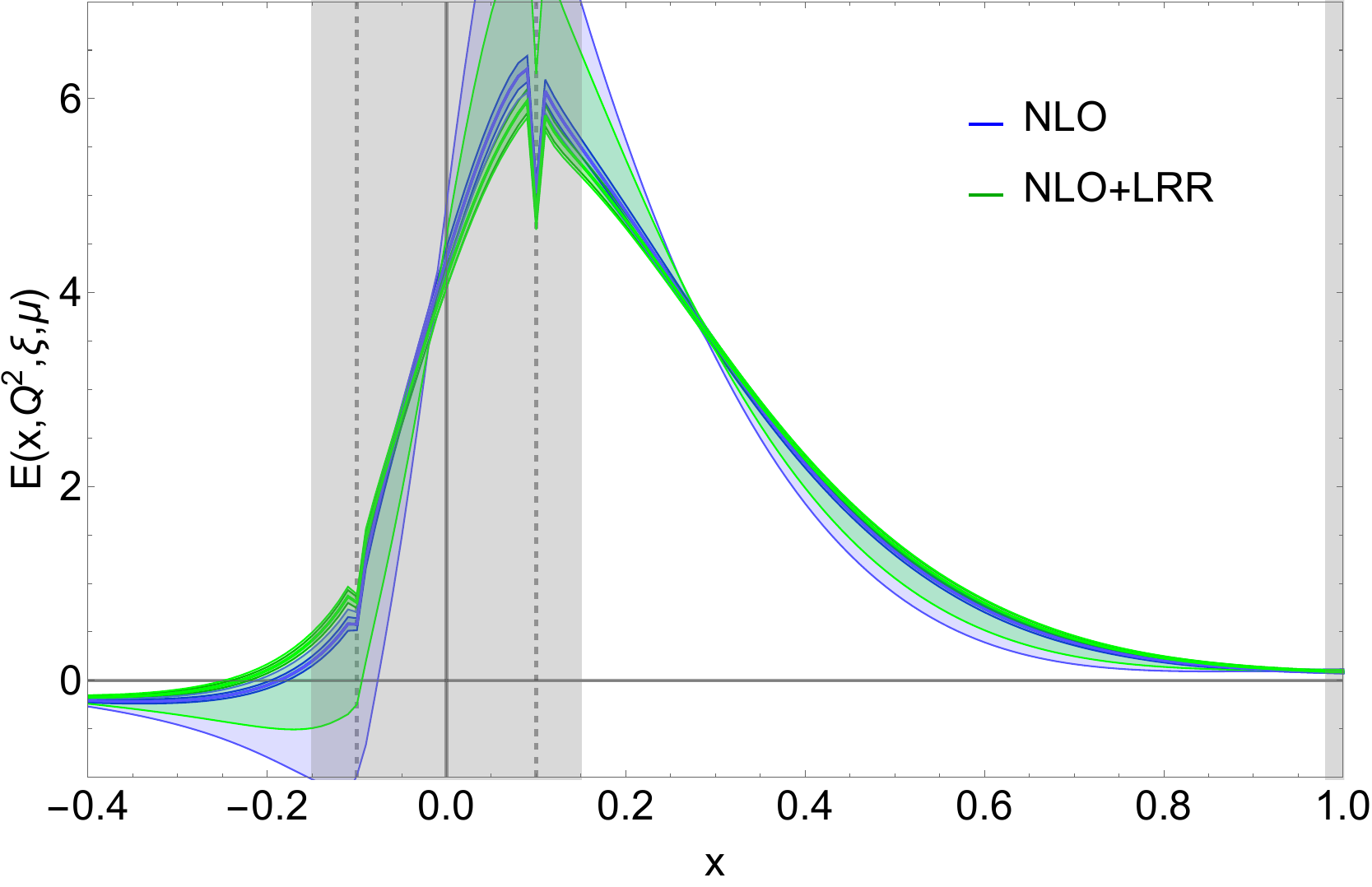}}
\caption{
Lightcone $H$ (left) and $E$ (right) GPDs evaluated at $\xi=0.1$ at \N\ (blue) and \NL\ (green).
The inner error bands are statistical and the outer error bands are combined statistical and systematic errors from scale variations.
The vertical dashed lines correspond to $x=\pm\xi$.
The GPDs suffer a discontinuity at these $x$ values due to the singularity in the matching kernel.
}
\label{fig:GPD-xi0p1}
\end{figure*}

The final unpolarized $H$ and $E$ GPDs are shown in Fig.~\ref{fig:GPD-xi0p1} for $\xi=0.1$ at \N\ and \NL;
we plot vertical dashed lines at $x=\pm\xi$.
As in the zero-skewness case, there is little change between the two aforementioned schemes in central values or error bars.
This is expected from the fact that the renormalon divergence has a lesser effect at fixed order (\N) than it does when RGR is included (\NR).
In addition, the GPD suffers a discontinuity at $x=\pm\xi$ due to the corresponding singularities in the matching kernel.
One difference between our nonzero-skewness $H$ GPD and those in Ref.~\cite{Alexandrou:2020zbe} is that our $H$ GPD does not plateau in the ERBL region.
The unpolarized $H$ GPD at $\xi=0.3$ in Fig.~3 of Ref.~\cite{Alexandrou:2020zbe} is approximately flat in the region $|x|<\xi=0.3$, whereas our $H$ GPD at $\xi=0.1$ increases in the region $|x|<\xi=0.1$.
Our ERBL region lies within the $x$-range where the LaMET expansion breaks down. For this reason, we should perhaps not expect our calculation to have the same qualitative behavior as that of Ref.~\cite{Alexandrou:2020zbe}.
The effect of LRR on the $x$-dependent GPD without RGR is similar to the corresponding effects at zero skewness (Fig.~\ref{fig:xi0HQ2-0});\
we, therefore, anticipate that the improvements we see for the unpolarized GPDs at zero skewness will also manifest at nonzero skewness once the methods have been adapted for the latter.
Because we are ultimately interested in the $x$ dependence, this is an auspicious indication of the benefits of RGR and LRR at $\xi\neq 0$.

\section{Conclusion and Outlook}\label{sec:Conclusion}

In this paper, we have shown the first application of leading-renormalon resummation and renormalization-group resummation to the unpolarized nucleon isovector GPD computed on the lattice in the framework of large-momentum effective theory.
We used a lattice spacing $a\approx 0.09$~fm with a physical pion mass, $N_f=2+1+1$ flavors of highly-improved staggered quarks and an average boost momentum $P_z\approx 2.2$~GeV with ensembles generated by the MILC collaboration~\cite{MILC:2010pul,MILC:2012znn,MILC:2015tqx}.
These matrix elements were renormalized in the hybrid-ratio scheme, applying RGR and LRR.
We then extrapolated the renormalized matrix elements to infinite distance and Fourier transformed to momentum space.
We report zero-skewness unpolarized nucleon GPDs, $H$ and $E$, with multiple momentum transfer values $Q^2$,
which have been matched to two loops as well as improved with both RGR and LRR for the first time.
The main advantage of the \NNLR\ calculation over other schemes is the reduction in systematic errors, since the central values remain compatible between the four schemes as shown in Sec.~2.
We also reported GPD functions $\xi=0.1$ at a single momentum transfer value of $Q^2=0.23$~GeV$^2$ in this work.
However, only the LRR improvement is applied to matching process up to one loop due to the lack an RGR calculation for nonzero-skewness GPDs to date.

The LaMET systematic errors were greatly reduced by the simultaneous application of RGR and LRR in the renormalization and matching processes.
For both the renormalized matrix elements and the $x$-dependent GPDs, the statistical errors remain approximately constant with the RGR and LRR modifications.
The improved systematics persist in the determination of the $x$-dependent GPDs.
The fact that systematic errors increase when we go from \N\ to \NN\ but decrease from \NLR\ to \NNLR\ show that the handling of systematics must keep pace with higher-order expansions in the matching and renormalization processes.
In addition, the systematic errors increased when RGR was applied on its own, due to its enhancement of the renormalon divergence.
The application of RGR and LRR to multiple $Q^2$ values at $\xi=0$ showed the efficacy of the two processes for nonzero momentum transfer.
Finally, we showed that the effects of RGR and LRR on the renormalized matrix elements and GPDs at nonzero skewness are also as promising as those at zero skewness.
Future work may involve the modification of the RGR matching to that of nonzero skewness using the ERBL equation in conjunction with the DGLAP equation.
In addition, the results could be further improved by performing the LaMET calculation at multiple boost momenta, $P_z$, in order to make an extrapolation $P_z\to\infty$ where the parton model is defined.

\section*{Acknowledgments}

We thank the MILC Collaboration for sharing the lattices used to perform this study.
JH thanks William Good for helpful discussions on the results of this work,
Yushan Su, Yong Zhao and Rui Zhang for discussions on LRR and RGR for the nonzero-skewness GPD,
and
Fei Yao and Jian-Hui Zhang for providing a notebook confirming the correct numerical implementation of the nonzero-skewness GPD matching kernel used for this work.
The LQCD calculations were performed using the Chroma software suite~\cite{Edwards:2004sx}.
This research used resources of the National Energy Research Scientific Computing Center (NERSC), a DOE Office of Science User Facility supported by the Office of Science of the U.S. Department of Energy under Contract No.~DE-AC02-05CH11231 through ERCAP;
Advanced Cyberinfrastructure Coordination Ecosystem: Services \& Support (ACCESS) program~\cite{boerner2023access}, which is supported by National Science Foundation grants 2138259, 2138286, 2138307, 2137603, and 2138296;
the Extreme Science and Engineering Discovery Environment (XSEDE)~\cite{towns2014xsede}, which was supported by National Science Foundation grant number 1548562;
facilities of the USQCD Collaboration, which are funded by the Office of Science of the U.S. Department of Energy,
and supported in part by Michigan State University through computational resources provided by the Institute for Cyber-Enabled Research (iCER).
The work of JH and HL are partially supported
by the US National Science Foundation under grant PHY 1653405 ``CAREER: Constraining Parton Distribution Functions for New-Physics Searches'', grant PHY~2209424,
by the U.S.~Department of Energy under contract DE-SC0024582,
and by the Research Corporation for Science Advancement through the Cottrell Scholar Award.

\bibliography{refs}
\end{document}